\documentclass{article}
\usepackage[latin1]{inputenc}
\usepackage[english]{babel}
\usepackage{dsfont}
\usepackage[cyr]{aeguill}
\usepackage{amsmath,amssymb}
\usepackage{amsthm}
\usepackage{eurosym}
\usepackage{multicol}
\usepackage{xcolor}
\usepackage{fancyhdr}
\usepackage[pdftex]{graphicx}
\usepackage{subcaption}
\usepackage[pdftex]{hyperref}
\usepackage{verbatim}
\usepackage{setspace}
\usepackage{stmaryrd}
\usepackage{amsmath,tikz}
\usetikzlibrary{matrix}

\newcommand{\indic}{\mathds{1}} 
\newcommand{\E}{\mathbb{E}} 
\DeclareMathOperator{\CD}{CD} 

\DeclareMathOperator{\argmin}{argmin}

\DeclareMathOperator{\Span}{span}

\theoremstyle{plain}
\newtheorem{thm}{Theorem}
\newtheorem{pro}{Proposition}

\providecommand{\keywords}[1]{\textbf{\textit{Keywords --}} #1}

\title{Prediction of linear fractional stable motions \\ using codifference, \\
with application to non-Gaussian rough volatility}

\author{Matthieu Garcin$^{\text{a,}}$\thanks{Corresponding author: matthieu.garcin@m4x.org. \newline 
$^{\text{a}}$ De Vinci Higher Education, De Vinci Research Center, Paris, France. \newline
$^{\text{b}}$ ESILV, 92916 Paris La D\'efense, France. \newline 
$^{\text{c}}$ Institute of Mathematics, Ecole Polytechnique F\'ed\'erale de Lausanne, Station 8, 1015 Lausanne, Switzerland. \newline
$^{\text{d}}$ Chair of Econophysics and Complex Systems, Ecole Polytechnique, 91128 Palaiseau, France. \newline
$^{\text{e}}$ LadHyX, UMR CNRS 7646, Ecole Polytechnique, Institut Polytechnique de Paris, 91128 Palaiseau, France. \newline
Acknowledgements: MG acknowledges the support of the Chair ``Deep Finance Statistics'' between QRT, Ecole Polytechnique, and its foundation. The authors are also grateful to Thomas Lux, Philippe Soulier, Gilles de Truchis, and to the participants of the 19th CFE-CMStatistics conference in London for useful discussions and comments.}, Karl Sawaya$^{\text{b,c}}$, Thomas Valade$^{\text{b,d,e}}$} 

\date{\today}

\begin{document}

\maketitle

\begin{abstract}
The linear fractional stable motion (LFSM) extends the fractional Brownian motion (fBm) by considering $\alpha$-stable increments. We propose a method to forecast future increments of the LFSM from past discrete-time observations, using the conditional expectation when $\alpha>1$ or a semimetric projection otherwise. It relies on the codifference, which describes the serial dependence of the process, instead of the covariance. Indeed, covariance is commonly used for predicting an fBm but it is infinite when $\alpha<2$. Some theoretical properties of the method and of its accuracy are studied and both a simulation study and an application to real volatility data, with a comparison to the fBm and to the heterogeneous auto-regressive model, confirm the relevance of the approach. The LFSM-based method shows a promising performance in the forecast of time series of volatilities, decomposing properly, in the fractal dynamic of rough volatilities, the contribution of the kurtosis of the increments and the contribution of their serial dependence. Moreover, the analysis of hit ratios suggests that, beside independence, persistence, and antipersistence, a fourth regime of serial dependence exists for fractional processes, characterized by a selective memory controlled by a few large increments.
\end{abstract}




\section{Introduction}

The fractional Brownian motion (fBm) extends the standard Brownian motion by introducing serial dependence between non-overlapping increments~\cite{MvN}. This property is crucial when it comes to forecast future increments~\cite{NP,BG}. The fBm is therefore popular in the financial industry to model log-prices, since such forecasts can be used to build systematic investment strategies~\cite{GNR,GMR,Garcin2017,Garcin2022,LAW}. It has also been used to forecast wind speed~\cite{AMN} or volatility~\cite{Garcin2022,BYZ}, making it useful for trading energy and weather derivatives or for developing volatility arbitrage strategies.

Beyond the interest of a serially dependent process, the empirical justification of the fBm lies in the reproduction of fractal properties of time series. However, serial dependence is not the only way to adjust specific fractal properties. The literature indeed mentions three complementary methods~\cite{CBMG}: the serial dependence, known as Joseph effect, the occurrence of large increments with a large probability~\cite{SamTaq}, known as Noah effect, and time-variation of the parameters~\cite{ST2006,BPP,Garcin2020}, known as Moses effect. It is possible to combine some of these three effects in various models~\cite{Garcin2020} but also to consider some other extensions to obtain additional properties, like the stationarity, which is useful for modelling rates or volatilities~\cite{CKM,HN,Viitasaari,Garcin2019,GarcinCNSNS}.

One can reproduce the Noah effect by introducing variables following a stable distribution instead of the Gaussian variables used in the definition of the standard Brownian motion and of the fBm. The thickness of the tails of the stable distribution is described by the stability parameter $\alpha\in(0,2]$. The smaller $\alpha$, the fatter the tails. The Gaussian distribution is a particular case of a stable distribution, with $\alpha=2$. Using this kind of stable variable leads to stable processes, like the stable L\'evy motion or the linear fractional stable motion (LFSM). The LFSM, which is the main model studied in this article, combines leptokurtic distribution and serial dependence. Stable dynamics are widely used to model phenomena in physics~\cite{SKR}, in telecommunication~\cite{CS,Cek}, in medicine~\cite{SGRS,SGK}, or in finance, both with the inclusion of a serial dependence~\cite{RD,Garcin2020,AG} and without~\cite{LL,BLVW,Nolan2003}. 

Methods for forecasting fractional stable processes would be very useful for practical applications and would extend the existing results of the fBm by properly separating the Noah and the Joseph effects. Indeed, applying the fBm methodology to leptokurtic time series would create a bias. Unfortunately, the extension of Gaussian forecasting methods to this class of distributions is not straightforward because the covariance, which is pivotal in the Gaussian approach~\cite{NP}, is infinite for stable variables as soon as $\alpha<2$. The purpose of this article is to propose a forecasting method for LFSM using tools other than covariance.

One can consider several alternatives to the covariance to build forecasting methods. First, though moments of order $p$ are infinite when $p\leq \alpha$, conditional moments can be finite for higher orders~\cite[Chapter 5]{SamTaq}, but it is under a set of assumptions which does not hold when future increments are obtained by adding independent variables to past observations, like in the LFSM~\cite[Theorem 5.1.3]{SamTaq}. The conditional expectation is however finite when another restrictive condition is met, namely when $\alpha>1$~\cite[Theorem 5.2.2]{SamTaq}. The obtained analytic formula can however not be generalized for a conditioning on a number of lagged increments larger than 1~\cite{ST91,HST}. A second alternative would consist in describing the dependence structure with copulas. It is however not convenient because it requires an expression for the cumulative distribution function, which, in the case of stable variables, can only be obtained numerically. A third alternative would be to depict the dependence by the mean of the multivariate characteristic function instead of the multivariate cumulative distribution function. Indeed, the spectral measure, which is derived from this multivariate characteristic function, entirely describes the dependence between stable variables. Since it is a function, the spectral measure is an object of infinite dimension, which is thus difficult to estimate from limited observations. Instead, one often considers the codifference~\cite{KT,BLVW,RD,WCG}, which simplifies the dependence structure contained in the spectral measure as the correlation or Kendall rank correlation do with the copula.

In this article, similarly to the Cholesky decomposition of a Gaussian vector, we propose a decomposition of discrete-time observations of an LFSM in independent stable variables. The decomposition has not exactly the same dependence structure as an LFSM but it has the same codifference. From this decomposition, we can propose a forecasting method. The evaluation of this method, first theoretically then on simulations, shows a new way of interpreting the parameters of an LFSM, namely the stability parameter $\alpha$ and the Hurst exponent $H$. Indeed, in addition to the traditional persistent, independent, and antipersistent cases, a new property of the time series appears for very small values of $\alpha$. The method is also compared to a previous study on the fBm using real time series of volatilities. The performance of our forecasting method in this real framework is promising and legitimises the model. Beyond the forecasting performance of the LFSM, we also highlight, through several methods, the non-Gaussian character of volatility series. We thus underline the relevance of a non-Gaussian version of rough volatility models, in which the LFSM can play a pivotal role. We also show that this model contains additional predictive information beyond that captured by the traditional heterogeneous autoregressive (HAR) volatility model.

There is an interesting, contemporary attempt to forecast another type of stable process, without the fractional feature, namely a discrete-time stable moving average~\cite{TFT}. Contrary to our contribution, it exploits the full dependence structure, that is the spectral measure, instead of the sole codifference. But it is intended to forecast the occurrence of extreme events only, whereas we propose a point forecast method.

The rest of the article is organized as follows. Section~\ref{sec:LFSM} presents the model along with some of its properties and the concept of codifference. In Section~\ref{sec:forecast_codiff}, we introduce the decomposition of  a vector of observations of the LFSM and deduce the forecasting method. A simulation study, with a comparison to the fBm and an autoregressive model, as well as an application to real data, namely time series of volatilities, are proposed respectively in Sections~\ref{sec:simul} and~\ref{sec:empirical}. Section~\ref{sec:conclusion} concludes.

\section{Model description: preliminaries on the LFSM}\label{sec:LFSM}

The LFSM is an extension of the fBm, with $\alpha$-stable increments instead of Gaussian ones~\cite{Taqqu87,KM,ST}. In the literature, the LFSM is also sometimes called fractional Lévy-stable motion~\cite{WBM} or fractional Lévy motion~\cite{Taqqu87,KM,WCG}. Since $\alpha$-stable variables do not have a finite variance as soon as $\alpha<2$, an alternative to autocovariance is to be used for quantifying the serial dependence of such a process. Our solution is based on codifference~\cite{KT}.

In this section, we first introduce codifference and LFSM, then we briefly present simulation and estimation methods.

\subsection{Codifference of jointly stable variables}\label{sec:DefPropCodiff}

Several equivalent definitions of a stable distribution exist~\cite{SamTaq}. Unfortunately, no analytic expression is available for the probability density function (pdf) but we know the characteristic function of an $\alpha$-stable random variable $X$, which is enough to define this distribution and from which we can get the pdf numerically thanks to an inverse Fourier transform~\cite{SGRS,AG}:
$$
\Phi_X:\theta\in\mathbb R \mapsto \mathbb{E}\left[e^{i \theta X}\right]=\left\{\begin{array}{ll} 
\exp \left(i \mu \theta-\gamma^{\alpha}|\theta|^{\alpha}\left[1-i \beta \frac{\theta}{| \theta |} \tan \left(\frac{\pi \alpha}{2}\right)\right]\right) & \text{if } \alpha \neq 1 \\ 
\exp \left(i \mu \theta-\gamma^{\alpha}|\theta|^{\alpha}\left[1+i \beta \frac{\theta}{| \theta |} \frac{2}{\pi} \ln|\theta|\right]\right) & \text{if } \alpha=1, 
\end{array}\right.
$$
where $\alpha \in (0,2]$ is the the stability parameter, $\beta \in [-1,1]$ is the skewness parameter, $\gamma \geq 0$, which can also be written $\Vert X\Vert_{\alpha}$, is the scale parameter, and $\mu \in \mathbb R$ is the location parameter.

In the particular cases where $\alpha =  2$ and $\alpha = 1$, we find respectively the characteristic function of a Gaussian distribution and of a Cauchy distribution. An $\alpha$-stable random variable is called symmetric $\alpha$-stable ($S\alpha S$) if it is symmetric, that is $\beta =\mu=0$.  In this case, the characteristic function reduces to 
\begin{equation}\label{eq:SaS}
\Phi_{X}(\theta) = e^{-\Vert X\Vert_{\alpha}^{\alpha} |\theta|^{\alpha}}.
\end{equation} 
In what follows, we will be dealing with $S\alpha S$ random variables, which will be said standard when $\Vert X\Vert_{\alpha}=1$.

We also note that an alternative parameterization of this distribution exists and is known as Zolotarev's (M) parameterization~\cite{Zolotarev}. This other parameterization avoids discontinuities of the probability density with respect to the parameters~\cite{Nolan}. However, when considering $S\alpha S$ variables, the two parameterizations coincide.

We are particularly interested in multivariate extensions of $S\alpha S$ variables and in the linear dependence between jointly stable variables, that is between the components of a stable vector~\cite[Definition 2.1.1]{SamTaq}. In general, this dependence cannot be described by the covariance, because the variance itself is infinite, unless $\alpha=2$. Indeed, when $\alpha<2$, $\E(|X|^p)<\infty$ for any $p\in(0,\alpha)$ and $\E(|X|^p)=\infty$ for any $p\geq\alpha$ \cite[Property 1.2.16]{SamTaq}. 

Several dependence measures are proposed in the literature to bypass this limitation~\cite[Chapter 2]{SamTaq}: covariation, L\'evy correlation cascade, codifference. Covariation has some practical limitations: it is restricted to $\alpha>1$ and it is based on the spectral measure which can be hardly retrieved with empirical data~\cite{Miller}. L\'evy correlation cascade exploits the Poisson process related to the LFSM~\cite{EK}. Our preference goes to codifference, which is valid whatever $\alpha\in(0,2]$ and which, among the three measures of dependence cited above, is the most related to the only available characterization of stable variables. It is indeed based on the characteristic function of random variables and it can be used for any type of probability distribution, and not only stable ones~\cite{KT,BLVW,RD,WCG}. 

The codifference between two random variables $X$ and $Y$ is defined by
$$\CD(X,Y) = -\ln(\Phi_X(1)) - \ln(\Phi_{Y}(-1)) + \ln(\Phi_{X-Y}(1)),$$
where $\Phi_Z$ is the characteristic function of any variable $Z$. In the case where $X$ and $Y$ are jointly $S\alpha S$ random variables, with $\alpha\in(0,2]$, the codifference more simply writes
$$\CD(X,Y) = \Vert X\Vert _{\alpha}^{\alpha} + \Vert Y\Vert _{\alpha}^{\alpha} - \Vert X-Y\Vert _{\alpha}^{\alpha},$$
after equation~\eqref{eq:SaS}. For jointly $S\alpha S$ random variables $X$ and $Y$, codifference also has some useful properties: it is symmetric in $X$ and $Y$; when $\alpha=2$ it is equal to covariance; it is equal to zero when $X$ and $Y$ are independent of each other, regardless of $\alpha\in(0,2]$, the converse being true when $\alpha\in(0,1)\cup\{2\}$~\cite[Properties 2.10.2-2.10.4]{SamTaq}.

The case of a linear combination of an arbitrary number of independent $S\alpha S$ random variables is of interest since it naturally appears when one builds a stochastic process with independent $S\alpha S$ increments. This is the purpose of Proposition~\ref{pro:sumSaS}. 

\begin{pro}\label{pro:sumSaS}
Let $d\in\mathbb N\setminus\{0\}$, $\alpha\in(0,2]$, $(a_1, ..., a_d)\in\mathbb R^d$, and $X_1,..., X_d$ be $d$ independent $S\alpha S$ random variables. Then $\sum_{i=1}^{d}a_iX_i$ is $S\alpha S$ and
\begin{equation}\label{eq:sumSaS}
\left\| \sum_{i=1}^{d}a_iX_i\right\|_{\alpha}^{\alpha} = \sum_{i=1}^{d}|a_i|^{\alpha}\left\|X_i\right\|_{\alpha}^{\alpha}.
\end{equation} 
\end{pro}

The proof of Proposition~\ref{pro:sumSaS} is postponed in Appendix~\ref{sec:proof_sumSaS}.

\subsection{The LFSM}\label{sec:DefPropLFSM}

The $S\alpha S$ L\'evy motion is the extension of a Brownian motion with an $S\alpha S$ distribution instead of the Gaussian one~\cite[Example 3.1.3]{SamTaq}. As well as the fBm extends the Brownian motion by modifying its serial dependence structure~\cite{MvN}, it is possible to extend the $S\alpha S$ L\'evy motion to reproduce specific fractal features. The obtained process is the LFSM $(X_t)_{t\in\mathbb{R}}$, of stability parameter $\alpha \in (0,2]$ and Hurst exponent $H\in(0,1)$, defined for all $t\in \mathbb{R}$ by
\begin{equation}\label{eq:flsm}
X_t = \int_{\mathbb R} \left[ (t-s)_+^{H-\frac{1}{\alpha}}-(-s)_+^{H-\frac{1}{\alpha}}\right]dL_{\alpha}(s),
\end{equation}
where $L_{\alpha}(s)$ is an $S\alpha S$ L\'evy motion~\cite{SamTaq,ST,WBM}.

The LFSM is thus a $H$-selfsimilar process with stationary $S\alpha S$ increments \cite[Proposition 7.4.2]{SamTaq}. The scale parameter of this process is also known to be as follows \cite[Proposition 7.4.3]{SamTaq}\cite{WCG}: 
\begin{equation}\label{eq:scalParamLFSM}
\forall t \geq 0 ,\ \Vert X_t\Vert _\alpha = K_{\alpha,H}|t|^{H},
\end{equation}
where 
$$K_{\alpha, H} = \left(\int_{\mathbb R}{\left|(1-s)_+^{H-\frac{1}{\alpha}} - (-s)_+^{H-\frac{1}{\alpha}}\right|^{\alpha}ds}\right)^{\frac{1}{\alpha}}.$$ 
The Hurst exponent reflects the dependence that exists between increments. If $H = 1/\alpha$, the increments of the LFSM are independent and the process is a standard $S\alpha S$ L\'evy motion. If $H > 1/\alpha$ (respectively $H < 1/\alpha$), the increments are positively (resp. negatively) dependent. The further the Hurst exponent is from $1/\alpha$, the strongest the serial dependence will be.  More precisely, the serial dependence structure of the LFSM can be described by its autocodifference~\cite{WCG}, which we obtain as a consequence of equation~\eqref{eq:scalParamLFSM} and of the stationarity of the increments of $X_t$:
\begin{equation}\label{eq:codiffLFSM}
\CD(X_t,X_s) = K_{\alpha,H}^{\alpha}\left(|t|^{H\alpha} + |s|^{H\alpha} -|t-s|^{H\alpha} \right).
\end{equation}
When $H=1/\alpha$ the above formula reduces to $\CD(X_t,X_s)=2\min(s,t)$.

\subsection{Simulation of the LFSM}

Historically, the first simulations of the fBm were based on a discretization of the integral definition of the fBm, that is equation~\eqref{eq:flsm} with $\alpha=2$~\cite{MvN}. This method only leads to an approximation of a true fBm, for two reasons: the integral is truncated by considering finite bounds, the continuous integral is replaced by a discrete sum. Beside this approximative method, many exact methods have been proposed. Most of these exact simulation methods of the fBm are based on the covariance matrix, like the Cholesky method, or the Davies-Harte and Wood-Chan ones~\cite{DH,Coeurjolly}. 

Unfortunately, these exact methods do not work for an LFSM when $\alpha<2$, because the covariance matrix is not defined. For this reason, the simulation of an LFSM is always approximative. The most popular approach is the Riemann-sum approximation of the stochastic integral representation of the LFSM of equation~\eqref{eq:flsm}~\cite[Section 7.11]{SamTaq}\cite{JW}, with the same two errors as in the fBm case cited above. It is also worth mentioning the existence of an approach that refines the Riemann-sum simulation method by the inclusion of a fast Fourier transform~\cite{CG,ST}.

\begin{figure}[h!]
      \centering
      \includegraphics[width=0.45\textwidth]{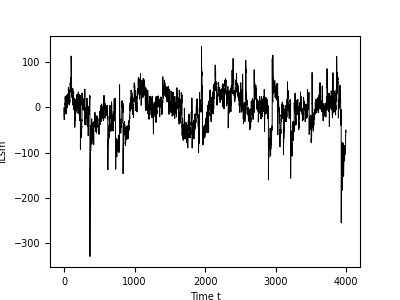}
      \includegraphics[width=0.45\textwidth]{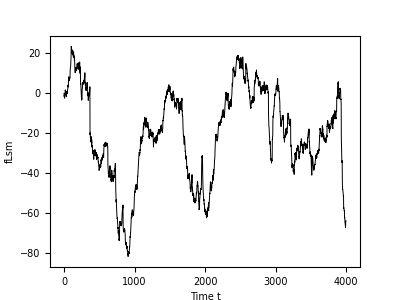} \\
      \includegraphics[width=0.45\textwidth]{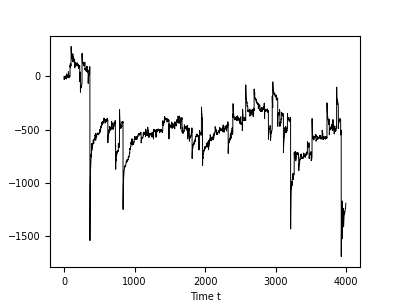}
      \includegraphics[width=0.45\textwidth]{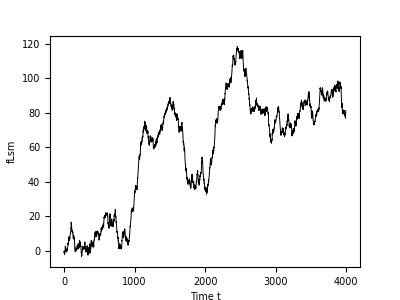}
\begin{minipage}{0.7\textwidth}\caption{Simulation of a path of a unitary LFSM, with the same seed used for the generation of pseudo-random variables, with $(\alpha,H) = (1.67,0.2)$ (top left), $(\alpha,H) = (1.67,0.8)$ (top right), $(\alpha,H) = (1.1,0.7)$ (bottom left), and $(\alpha,H) = (1.95,0.7)$ (bottom right). The time step is 0.01.}
\label{fig:simul}
\end{minipage}
\end{figure}

We thus discretize the integral with a ``small'' time step, equal to 0.01 in the examples displayed in Figure~\ref{fig:simul}. We then simply calculate the deterministic integrand for each time interval and multiply it by a random $S\alpha S$ variable. All the $S\alpha S$ variables are independent of each other. We simulate them using the Chambers-Mallows-Stuck method~\cite{CMS,Weron,SGK}: first we simulate two independent variables, $P$ uniform in $(-\pi/2,\pi/2)$ and $Q$ exponential of parameter 1, then we combine them with the following formula to get the unitary $S\alpha S$ variable $R$:
$$R=\frac{\sin(\alpha P)}{\left(\cos(P)\right)^{1/\alpha}}\left(\frac{\cos\left(P(1-\alpha)\right)}{Q}\right)^{\frac{1-\alpha}{\alpha}}.$$
We note that this method is not valid when $\alpha= 1$, but this case corresponds to the Cauchy distribution, whose cumulative distribution function is explicitly known and can be used, after inversion, for simulation purposes.

In the simulations  shown in Figure~\ref{fig:simul}, we have two processes with a negative serial dependence, that is a negative autocodifference, for $H=0.2$ (and $\alpha=1.67$) but also for a value of $H$ which would be associated to positive autocorrelation in the case of an fBm, namely $(\alpha,H) = (1.1,0.7)$.

\subsection{Estimation of the LFSM}\label{sec:estim}

Several approaches are proposed in the literature to estimate the two parameters $\alpha$ and $H$ of the LFSM. Estimators based on the wavelet transform have appealing asymptotic properties and are used for the standalone estimation of either $H$~\cite{SPT,PTA2007} or $\alpha$~\cite{AH}. Alternative joint estimation methods can be based on power variations~\cite{GLT,Shergin}, which supposes that one selects a power lower than a known lower bound of $\alpha$. Using an empirical characteristic function is also a natural choice for designing an estimator of both $H$ and $\alpha$~\cite{MOP,LP}. It indeed follows a widespread method used for estimating the $\alpha$ parameter of a stable distribution~\cite{Koutrouvelis,KW,SGK}, among other methods~\cite{Nolan2020,AG}, like the one based on empirical quantiles~\cite{McCulloch}.

In this work, the codifference, which is based on the characteristic function, is a central concept, so we use estimators based on the empirical characteristic function. We could also use empirical codifferences~\cite{WCG}, which is an aggregation of empirical characteristic functions.

Let $(X_t)_{t\in\mathbb R}$ be an LFSM. Following Section~\ref{sec:DefPropLFSM}, we have, for all $\theta\in\mathbb R$:
\begin{equation}\label{eq:estim_par_charac}
\ln\left(\Phi_{X_{.+\tau}-X_.}(\theta)\right)=-K_{\alpha,H}^{\alpha}|\tau|^{\alpha H}|\theta|^{\alpha}.
\end{equation}
Fixing alternatively $\tau$ and $\theta$, one can successively estimate $\alpha$ and $H$. 

One starts by selecting a reference time step, $\tau_0$. It can for example be the smallest time step in the dataset, so that one can count on numerous observations, unless the dataset is disrupted by a microstructure noise, like a truncation of the numbers as it appears for prices in financial markets. It is indeed well known that such a noise may lead to a biased estimation of the selfsimilarity parameter, with a stronger effect for higher frequencies~\cite{LSG,GG,GarcinCNSNS}. Then, we focus on the linear regression of $\ln(-\ln(\widehat \Phi_{X_{.+\tau_0}-X_.}(\theta)))$ on $\ln|\theta|$, for a well-chosen set of values of $\theta$, where $\widehat \Phi_Y$ is the real part of the empirical characteristic function of $Y$, of which we observe $n$ replications $Y_1$, ..., $Y_n$:
$$\widehat \Phi_Y:\theta\longmapsto\frac{1}{n}\sum_{i=1}^{n}\cos(\theta Y_i).$$
In the simulation study, Section~\ref{sec:simul}, the set of values for $\theta$ is $\{1,2,...,20\}$. If one considers much smaller values for $\theta$, we have a problem of identifiability since the cosine is very close to 1, whatever $\alpha$~\cite{SGK}. The slope $S_1$ of the above linear regression must converge to $\alpha$, after equation~\eqref{eq:estim_par_charac}.

Similarly, by fixing $\theta=1$, which is the most natural value for $\theta$ when one is interested in codifference, we consider the linear regression of $\ln(-\ln(\Phi_{X_{.+\tau}-X_.}(1)))$ on $\ln|\tau|$, whose slope $S_2$ must converge to $\alpha H$.

Finally, the estimators of $\alpha$ and $H$ are:
$$\left\{\begin{array}{ccl}
\widehat{\alpha} & = & S_1 \\
\widehat{H} & = & S_2/S_1.
\end{array}\right.$$

\begin{figure}[htbp]
	\centering
		\includegraphics[width=0.45\textwidth]{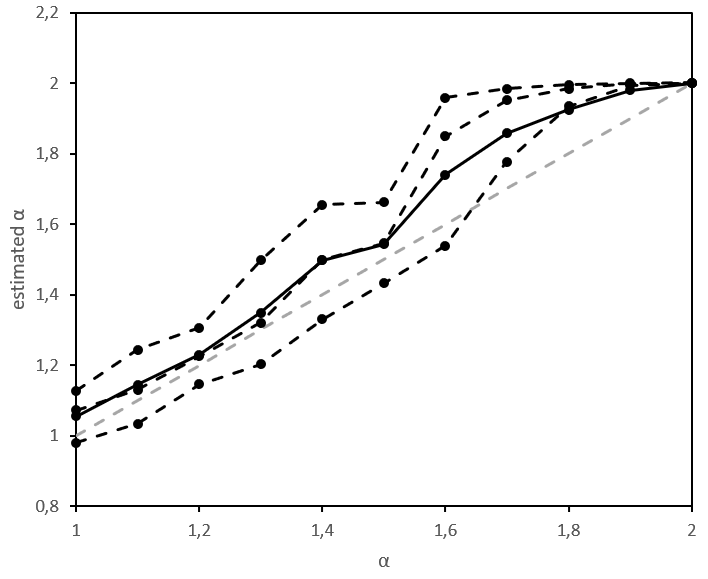} 
		\includegraphics[width=0.45\textwidth]{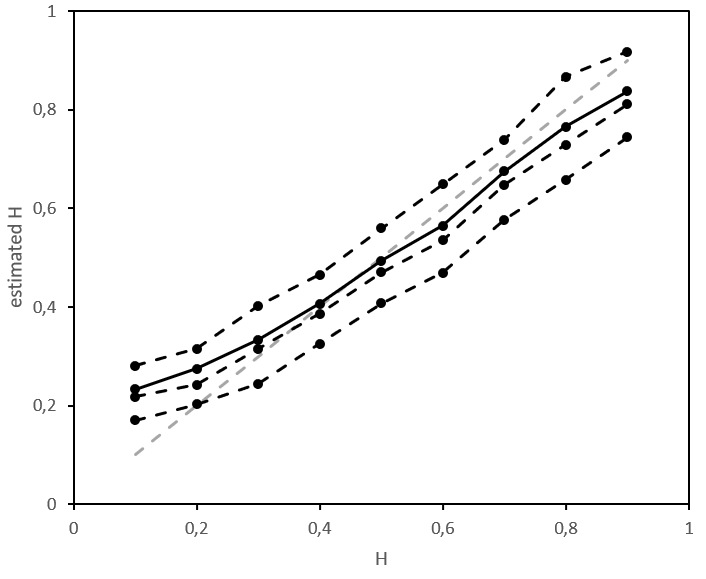}
\begin{minipage}{0.7\textwidth}\caption{Estimation of $\alpha$ (left) and $H$ (right) for an LFSM with $\alpha\in[1,2]$ and $H=0.8$ (left) or $\alpha=1.5$ and $H\in[0.1,0.9]$ (right). The solid line is the average estimate, the black dotted lines are quantiles of probability $25\%$, $50\%$, and $75\%$ of the estimates. The grey line is the ideal value.}
	\label{fig:estim}
\end{minipage}
\end{figure}

We show in Figure~\ref{fig:estim} the output of this estimation method for a simulated LFSM. For a fixed pair $(\alpha,H)$, we simulate 100 trajectories in the time interval $[0,10]$ with a time step 0.01. The simulation, based on the integral definition, is an approximation of an LFSM. So, to restrict the effects of this approximation, we use a larger time scale in the estimator, with $\tau_0=0.1$. We estimate $\alpha$ and $H$ for each trajectory and represent in Figure~\ref{fig:estim} the average and the quartiles of the 100 estimates. We note that we focus on $\alpha\in[1,2]$ because this interval contains all the estimated values of $\alpha$ in our financial dataset, Section~\ref{sec:empirical}.

The estimation method detailed above stands for a standard LFSM, that is with a scale parameter equal to $K_{\alpha,H}$ for an increment of duration 1. But there is no reason to have such a property for real time series such as those of Section~\ref{sec:empirical}. Instead, the time series will be modelled by $\sigma X_t$, with $\sigma>0$ and $X_t$ an LFSM. In our empirical study, where $\alpha >1$, we obtain, from the formula of absolute moments given in Section~\ref{sec:quality}, that
$$\E\left(\left|X_{.+\tau_0}-X_.\right|\right)=\frac{2\Gamma(1-\alpha^{-1})}{\pi}\sigma.$$
Replacing the expectation by its empirical counterpart and plugging $\widehat{\alpha}$ in the above equation, we get a straightforward estimator for $\sigma$.

\section{Forecast of LFSM with codifference}\label{sec:forecast_codiff}

This section introduces first a decomposition of a stable process in discrete time. Then, we present how this decomposition is to be used to forecast a future value of the process. Last, we evaluate theoretically the quality of this forecast.

\subsection{Discrete-time decomposition of stable processes}\label{sec:decompo}

When working with Gaussian processes, even with the fBm, the mean and covariance matrix are enough to describe the distribution of a vector of discrete-time observations of this process. This allows, for example, to decompose the components of this vector into sums of independent Gaussian variables for simulation purposes~\cite{Coeurjolly}, or to forecast future values by a conditional expectation, which we obtain by manipulating matrices~\cite{NP,Garcin2022}. Unfortunately, the stable non-Gaussian case is not a straightforward extension. It has for example been proved that non-trivial continuous-time stable processes do not admit a Karhunen-Loève decomposition, unless $\alpha=2$~\cite{PTA}. The codifference, introduced in Section~\ref{sec:DefPropCodiff}, is in fact not enough to describe all the dependence structure of a stable process. Worse, even the dependence structure of a simple jointly $S\alpha S$ finite-dimensional vector is not totally described by its codifference matrix, that is the matrix containing the codifference between all pairs of components. The missing piece, which totally characterizes the dependence structure, is the spectral measure, which is a measure on the unit sphere $\mathbb S^{d-1}$ of $\mathbb R^d$~\cite[Section 2.3]{SamTaq}. It appears in the characteristic function of an $S\alpha S$ vector $\mathbf X=(X_1,...,X_d)'$,
$$\Phi_{\mathbf X}(\boldsymbol{\theta})=\E\left[e^{i\langle\boldsymbol{\theta},\mathbf X\rangle}\right]=\exp\left(-\int_{\mathbb S^{d-1}}|\langle\boldsymbol{\theta},\mathbf{s}\rangle|^{\alpha}\Gamma_{\mathbf X}(d\mathbf s)\right),$$
where $\Gamma_{\mathbf X}$ is the spectral measure, $\boldsymbol{\theta}\in\mathbb R^d$, and $\langle.,.\rangle$ is the scalar product in $\mathbb R^d$. The codifference between two $S\alpha S$ variables only summarizes in a scalar the dependence structure contained in the spectral measure:
$$\CD(X_1,X_2)=\int_{\mathbb S^{1}}\left(|s_1|^{\alpha}+|s_2|^{\alpha}-|s_1-s_2|^{\alpha}\right)\Gamma_{(X_1,X_2)'}(d\mathbf s).$$
Therefore, two distinct stable vectors, that is with two distinct spectral measures, may have the same codifference matrix. The limitation of the codifference to characterize multivariate stable distributions can also be seen in Section~\ref{sec:DefPropCodiff}, where we saw  that a zero codifference is a necessary but not sufficient condition for independence.

Since the purpose of this work is to build a forecasting method exploiting the sole codifference, we introduce a transformation of a continuous-time $S\alpha S$ process $(X_t)_{t\in\mathbb R}$ in a discrete-time process $(\mathcal T X_t)_{t\in\mathbb Z}$, such that, whatever $t\in\mathbb Z$, the vector $(X_t,...,X_{t+d-1})'$ has the same $d\times d$ codifference matrix as $(\mathcal T X_t,...,\mathcal T X_{t+d-1})'$ and, $\forall i\in\llbracket 0,d-1\rrbracket$ and $\forall t\in\mathbb Z$,
\begin{equation}\label{eq:transfodiscr}
\mathcal T X_{t+i}=\sum_{j=0}^{i} a_{t,i,j} Z_j,
\end{equation}
where $(Z_0,...,Z_{d-1})'$ is a vector of jointly $S\alpha S$ independent variables of unitary scale parameter, $a_{t,i,j}\in\mathbb R$ for all $i,j\in\llbracket 0,d-1\rrbracket$, and where we note for convenience $a_{t,i,j}=0$ when $j>i$. We don't study this decomposition in the general case and we focus on the LFSM. Theorem~\ref{thm:systDecompo} shows that, under a limited set of assumptions, if we find such a decomposition for the LFSM, this decomposition is unique. 

\begin{thm}\label{thm:systDecompo}
Let $(X_t)_{t\in\mathbb R}$ be an LFSM and $t\in\mathbb N$. Whatever $d\in\mathbb N\setminus\{0\}$, if the decomposition proposed in equation~\eqref{eq:transfodiscr} exists, with $(X_t,...,X_{t+d-1})'$ having the same codifference matrix as $(\mathcal T X_t,...,\mathcal T X_{t+d-1})'$, and if the coefficients $a_{t,i,j}$ verify the condition that $a_{t,i,j}> 0$ for $i\geq j$ and $a_{t,\ell,j}> a_{t,i,j}$ (respectively $a_{t,\ell,j}< a_{t,i,j}$ and $a_{t,\ell,j}=a_{t,i,j}$) for $j\leq i\leq \ell$ when $H>1/\alpha$ (resp. $H<1/\alpha$ and $H=1/\alpha$), then the decomposition is unique and is the solution of the following system of $d(d+1)/2$ nonlinear equations:
$$\left\{\begin{array}{ll}
(\mathcal E_{i,i}) & |a_{t,i,i}|^{\alpha} = K_{\alpha,H}^{\alpha}|t+i|^{\alpha H} - \sum_{j=0}^{i-1} |a_{t,i,j}|^{\alpha} \\
(\mathcal E_{\ell,i}) & f_{t,i}(a_{t,\ell,i}) = K_{\alpha,H}^{\alpha}\left(|t+\ell|^{\alpha H}-|\ell-i|^{\alpha H}\right) - \sum_{j=0}^{i-1} \left(|a_{t,\ell,j}|^{\alpha} - |a_{t,\ell,j}-a_{t,i,j}|^{\alpha}\right)
\end{array}\right.$$
with equation $(\mathcal E_{i,i})$ defined for $i\in\llbracket 0,d-1\rrbracket$, equation $(\mathcal E_{\ell,i})$ defined for $(i,\ell)\in\llbracket 0,d-2\rrbracket\times\llbracket i+1,d-1\rrbracket$, and $f_{t,i}(z)=|z|^{\alpha}- |z-a_{t,i,i}|^{\alpha}$.
\end{thm}

The proof of Theorem~\ref{thm:systDecompo} is postponed in Appendix~\ref{sec:proof_systDecompo}.

From the system of nonlinear equations provided in Theorem~\ref{thm:systDecompo}, we propose a simple algorithm to determine the coefficients $a_{t,i,j}$ of the decomposition~\eqref{eq:transfodiscr}, when $H$ and $\alpha$ are known. We solve the equations $(\mathcal E_{i,j})$ one after the other, following the lexicographical order: $(\mathcal E_{0,0})$ first, then $(\mathcal E_{1,0})$, $(\mathcal E_{1,1})$, $(\mathcal E_{2,0})$, $(\mathcal E_{2,1})$, $(\mathcal E_{2,2})$, ..., $(\mathcal E_{d-1,d-1})$. For each equation $(\mathcal E_{i,j})$, we use the coefficients obtained at the previous steps and lexicographically ordered before $a_{t,i,j}$. When $i=j$, a straightforward solution to $(\mathcal E_{i,j})$ is available. When $i\neq j$, we obtain a numerical solution very rapidly for $a_{t,i,j}$, with a Newton-Raphson algorithm initiated at a value slightly higher (respectively lower) than $a_{t,i-1,j}$ when $H>1/\alpha$ (resp. $H<1/\alpha$). As one can see in the proof of Theorem~\ref{thm:systDecompo}, the case $H=1/\alpha$ is simpler and does not require any numerical optimization since, in this case, we have 
$$a_{t,i,j}=t^{\frac{1}{\alpha}\indic_{j=0}}\indic_{i\geq j}.$$
The case $\alpha=2$ is also simple and does not require any optimization since $f_{t,i}(a_{t,\ell,i})=2a_{t,\ell,i}a_{t,i,i}-a_{t,i,i}^2$, making it possible to isolate $a_{t,\ell,i}$ in $(\mathcal E_{\ell,i})$. This is consistent with the decomposition and forecasting framework for the fBm, which is already well known and only based on matrix manipulations~\cite{NP,Garcin2022}.

\begin{figure}[htbp]
	\centering
		\includegraphics[width=0.45\textwidth]{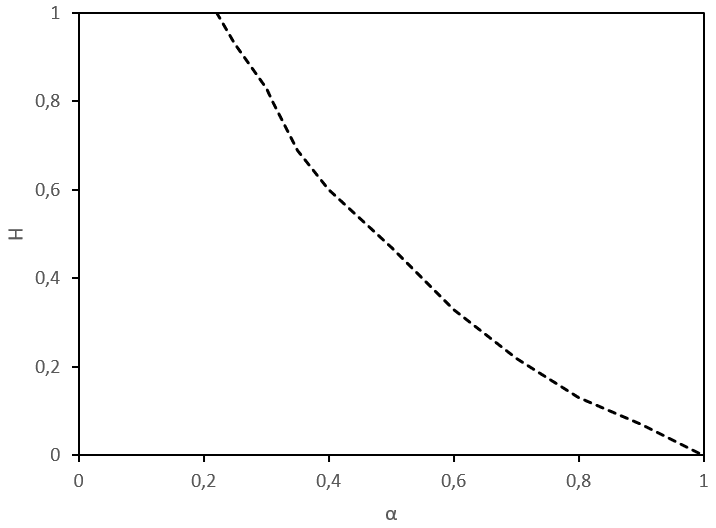} 
		\includegraphics[width=0.45\textwidth]{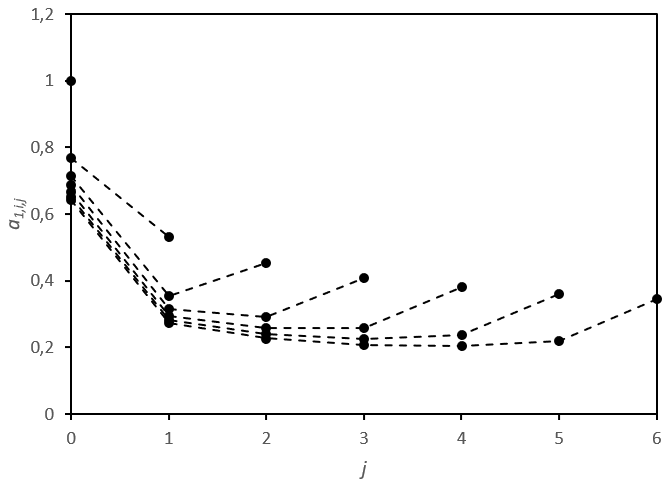} \\
		\includegraphics[width=0.45\textwidth]{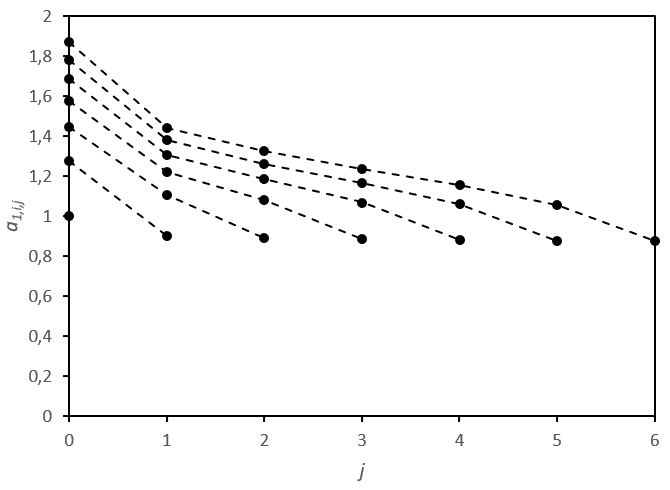} 
		\includegraphics[width=0.45\textwidth]{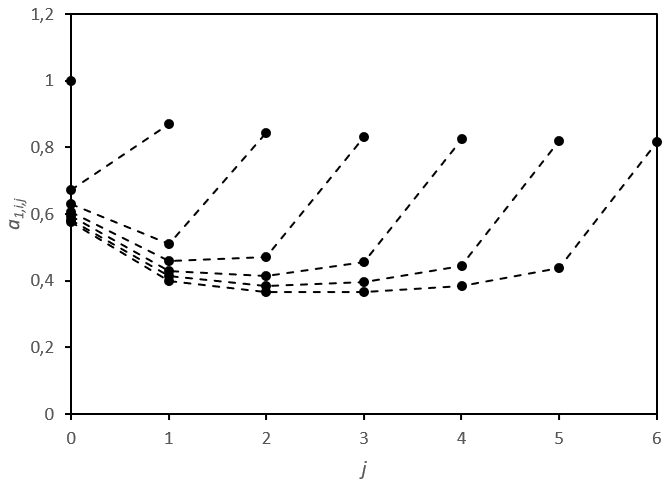} 
\begin{minipage}{0.7\textwidth}\caption{Top left: Frontier of the pairs $(\alpha,H)$ above which we numerically get the existence of the coefficients $a_{t,i,j}$ respecting the constraints of Theorem~\ref{thm:systDecompo}, for $t=1$ and $i\in\llbracket 0,6\rrbracket$. Three other graphs: coefficients $a_{1,i,j}$ for $i$ between 0 and 6 and $j\in\llbracket 0,i\rrbracket$ (one curve for each $i$), with $(\alpha,H)$ successively equal to $(0.7,0.8)$ (top right), $(1.5,0.8)$ (bottom left), and $(1.5,0.3)$ (bottom right).}
	\label{fig:coefDecompo}
\end{minipage}
\end{figure}

Theorem~\ref{thm:systDecompo} focuses on the uniqueness of the solution but not on its existence. However, we have conducted numerical tests with many values of $H$ and $\alpha$ and have always found a solution fulfilling the conditions of the theorem, except for very small values of $\alpha$, much smaller than the ones estimated in the empirical study of Section~\ref{sec:empirical}, as displayed in Figure~\ref{fig:coefDecompo}. It is also possible to prove theoretically the existence of the coefficients for small values of $i$. For example, assuming $H>1/\alpha$, in order to prove the existence of $a_{t,1,0}$, we look for $z>a_{t,0,0}=K_{\alpha,H}|t|^{H}$ such that $f_{t,0}(z)=K_{\alpha,H}^{\alpha}(|t+1|^{\alpha H} -1)$. We note that $f_{t,0}(a_{t,0,0})=K_{\alpha,H}^{\alpha}|t|^{\alpha H}$. The mapping $g:x\mapsto x^{\alpha H}$ is convex because $H>1/\alpha$ and $g(0)=0$, so $g$ is superadditive. Therefore $f_{t,0}(a_{t,0,0})\leq K_{\alpha,H}^{\alpha}(|t+1|^{\alpha H} -1)$. We also note that $f_{t,0}(K_{\alpha,H}|t+1|^{H})=K_{\alpha,H}^{\alpha}(|t+1|^{\alpha H}-[(t+1)^H-t^H]^{\alpha})$. Since $H\in(0,1)$, the mapping $h:x\mapsto x^H$ is concave and $h(0)=0$ so it is subadditive. Therefore $(t+1)^H\leq t^H+1$ and $f_{t,0}(K_{\alpha,H}|t+1|^{H})\geq K_{\alpha,H}^{\alpha}(|t+1|^{\alpha H} -1)$. Finally, by continuity of $f_{t,0}$, we know that a solution $a_{t,1,0}$ to equation $(\mathcal E_{1,0})$ exists and that $a_{t,1,0}\in\left[a_{t,0,0},K_{\alpha,H}|t+1|^{H}\right]$.

The coefficients in equation~\eqref{eq:transfodiscr} depend on $t$. For example, the coefficients $(a_{1,i,j})_{i,j\in\llbracket 0,d-1\rrbracket}$ are to be used for the vector $(\mathcal T X_1,...,\mathcal T X_{d})'$. But, in order not to calculate again the coefficients, if one wants the same kind of decomposition later in the time series, say for $(X_{t},...,X_{t+d-1})'$, one can define the translated process $Y_s=X_{t-1+s}-X_{t-1}$, which is also an LFSM since $Y_0=0$. Then we can apply Theorem~\ref{thm:systDecompo} to $(\mathcal T Y_1,...,\mathcal T Y_{d})'$ and use the same coefficients $(a_{1,i,j})_{i,j\in\llbracket 0,d-1\rrbracket}$.

\subsection{From the decomposition to the forecast}\label{sec:Predictor}

By exploiting the codifference-based decomposition of an $S\alpha S$ process $X_t$ in a sum of iid $S\alpha S$ variables, as introduced in equation~\eqref{eq:transfodiscr}, one can easily build methods to forecast the process $X_t$ at a future time. The simplest solution is a conditional expectation but we will see that it is restricted to the case $\alpha>1$. Therefore, we will have to introduce another method, based on a metric projection, and exploiting also equation~\eqref{eq:transfodiscr}.

We assume we observe the process at $d-1$ discrete times: $X_1,...,X_{d-1}$. We want to forecast $\mathcal T X_d$, where we define $\mathcal T X_t$ like in Section~\ref{sec:decompo} as a linear combination of independent $S\alpha S$ variables and such that $\mathcal T X_t=X_t$ for all $t\in\llbracket 1,d-1\rrbracket$. As exposed in Section~\ref{sec:decompo}, it can be extended to cases where all the times are translated by $t$. 

\subsubsection{Conditional expectation: $\alpha\in(1,2]$}\label{sec:condExp}

We assume that $\mathbf X_{1,d}=(\mathcal T X_1,...,\mathcal T X_{d})'$ follows
\begin{equation}\label{eq:decomposMatrix}
\mathbf X_{1,d}=\mathbf A_{0,d-1} \mathbf Z_{0,d-1},
\end{equation}
where $\mathbf Z_{0,d-1}=(Z_0,...,Z_{d-1})'$ is a jointly $S\alpha S$ independent vector of unitary scale parameter and the matrix $\mathbf A_{0,d-1}\in\mathbb R^{d\times d}$, of element $[\mathbf A_{0,d-1}]_{ij}=a_{1,i-1,j-1}$, is defined in accordance with the details given in Section~\ref{sec:decompo}. 

Inspired by the Gaussian case~\cite{NP}, one can be tempted to build a forecast of $X_d$ by considering its conditional expectation. For $S\alpha S$ variables, the conditional expectation is well defined for $\alpha>1$. When $\alpha\leq 1$, it can also be well defined under some restrictive conditions~\cite[Chapter 5]{SamTaq}. In our case, since $\mathcal T X_d$ is obtained by adding an independent increment to a linear combination of $\mathcal T X_1$, ..., $\mathcal T X_{d-1}$, the conditional expectation $\E[\mathcal T X_d|\mathbf X_{1,d-1}]$ only exists for $\alpha>1$. In this case, by linearity of the expectation, equation~\eqref{eq:decomposMatrix} and $\mathbf X_{1,d-1}=\mathbf A_{0,d-2} \mathbf Z_{0,d-2}$ give the following forecast:
\begin{equation}\label{eq:condExp}
\widehat{\mathcal T X}_{d}=\E\left[\mathcal T X_d\left|\mathbf X_{1,d-1}\right.\right] = \sum_{j=0}^{d-2} a_{1,d-1,j}Z_j.
\end{equation}

We now summarize the steps of the cascade algorithm, which leads to the forecast $\widehat{\mathcal T X}_{d}$. For $i$ successively equal to 0, ..., $d-2$, we do the following: 
\begin{enumerate}
\item[Step 1:] calculate $a_{1,i,0}$, ..., $a_{1,i,i}$, solving successively $(\mathcal E_{i,0})$, ..., $(\mathcal E_{i,i})$,
\item[Step 2:] determine $Z_i$, defined as $(X_{i+1}-\sum_{j=0}^{i-1}a_{1,i,j}Z_j)/a_{1,i,i}$.
\end{enumerate}
At the end, we also compute step 1 for $i=d-1$, so that we can calculate $\widehat{\mathcal T X}_{d}$ using equation~\eqref{eq:condExp}. 

In step 2, we use $X_{i+1}$, which is observed. When working conditionally on past observations $X_1$, ..., $X_d$, we assume that $\mathcal T X_i=X_i$ for $i\leq d-1$. We may only have a divergence between $\mathcal T X_d$ and $X_d$, which are both unobserved at this date. In other words, we forecast $\mathcal T X_d$ based on a model that might be slightly different from the one of $X_d$,\footnote{ They only have the same codifference, not necessarily the same spectral measure.} but using exactly the same conditioning on past observations.

\subsubsection{Metric and semimetric projections: $\alpha\in(0,2]$}

Let's consider the space $V_{j,k}=\Span(Z_j,...,Z_k)$, where the variables $Z_0$, ..., $Z_{d-1}$, derived from $\mathcal T X_1$, ..., $\mathcal T X_d$, are those introduced in equation~\eqref{eq:transfodiscr}. When $\alpha=2$, the justification of the conditional expectation as a predictor relies on the fact that it is an orthogonal projection of $\mathcal T X_d$ onto the space $V_{0,d-2}$, with the covariance as inner product. For other values of $\alpha$, the covariance is not defined and this approach cannot directly be extended to the codifference because, as being not bilinear, it cannot be an inner product. However, replacing the orthogonal projection by another kind of projection, we can still use the above decomposition in a sum of independent $S\alpha S$ variables to build a predictor.
 
Considering a variable $Y\in V_{0,d-1}$, its metric projection onto $V_{0,d-2}$ is the variable $Z\in V_{0,d-2}$ that minimizes a certain metric $D$. The function $D:V_{0,d-1}\times V_{0,d-1}\longrightarrow \mathbb R$ is a metric if, $\forall U,W,Y\in V_{0,d-1}$,
\begin{enumerate}
\item[(i)] $D(U,W)\geq 0$, with equality iff $U=W$,
\item[(ii)] $D(U,W)=D(W,U)$,
\item[(iii)] $D(U,W)\leq D(U,Y)+D(Y,W)$.
\end{enumerate}
In particular, $D(U,W)=\Vert U-W\Vert_{\alpha}$ is a metric if $\alpha\in[1,2]$. Indeed, if $U$ and $W$ respectively write $\sum_{i=0}^{d-1} \gamma^U_i Z_i$ and $\sum_{j=0}^{d-1} \gamma^W_j Z_j$, by independence of the $Z_j$ and Proposition~\ref{pro:sumSaS}, we have $D(U,W)=(\sum_{j=0}^{d-1}|\gamma^U_j-\gamma^W_j|^{\alpha})^{1/\alpha}$. Conditions (i) and (ii) are clearly satisfied, but condition (iii) only holds when $\alpha\geq 1$, after Minkowski inequality. In the case $\alpha\in(0,1)$, the triangle inequality is missing and $D$ is only a semimetric~\cite{Wilson}. Theorem~\ref{thm:projMetric} shows that the metric or semimetric projection leads to a unique predictor, $\widehat{\mathcal T X}^{D,V_{0,d-2}}_{d}$ which is the same as the one obtained in Section~\ref{sec:condExp}: $\widehat{\mathcal T X}^{D,V_{0,d-2}}_{d}=\widehat{\mathcal T X}_d$.

\begin{thm}\label{thm:projMetric}
Let $(X_t)_{t\in\mathbb R}$ be an LFSM. Whatever $d\in\mathbb N\setminus\{0,1\}$, if the decomposition proposed in equation~\eqref{eq:transfodiscr} exists, the metric (or semimetric if $\alpha\in(0,1)$) projection from $V_{0,d-1}$ onto $V_{0,d-2}$,
$$\widehat{\mathcal T X}^{D,V_{0,d-2}}_{d}=\underset{U\in V_{0,d-2}}{\argmin} D(\mathcal T X_d,U),$$
is unique and such that
$$\widehat{\mathcal T X}^{D,V_{0,d-2}}_{d}= \sum_{j=0}^{d-2} a_{1,d-1,j}Z_j,$$
where the coefficients $a_{1,d-1,j}$ are those of equation~\eqref{eq:transfodiscr}.
\end{thm}

The proof of Theorem~\ref{thm:projMetric} is postponed in Appendix~\ref{sec:proof_projMetric}.

While the codifference is not an inner product of $V_{0,d-1}$, we can see that the residual $\mathcal T X_d - \widehat{\mathcal T X}^{D,V_{0,d-2}}_{d}$ of the above projection has a zero codifference with any element of $V_{0,d-2}$. Indeed, the residual also writes $a_{1,d-1,d-1}Z_{d-1}$, after equation~\eqref{eq:transfodiscr} and Theorem~\ref{thm:projMetric}, and its codifference with $\sum_{j=0}^{d-2}\gamma_jZ_j\in V_{0,d-2}$ is
$$\begin{array}{cl}
 & \CD\left(\mathcal T X_d - \widehat{\mathcal T X}^{D,V_{0,d-2}}_{d},\sum_{j=0}^{d-2}\gamma_jZ_j\right) \\
 = & \Vert a_{1,d-1,d-1}Z_{d-1}\Vert_{\alpha}^{\alpha} + \Vert \sum_{j=0}^{d-2}\gamma_jZ_j\Vert_{\alpha}^{\alpha} - \Vert a_{1,d-1,d-1}Z_{d-1}-\sum_{j=0}^{d-2}\gamma_jZ_j\Vert_{\alpha}^{\alpha} \\
 = & |a_{1,d-1,d-1}|^{\alpha} + \sum_{j=0}^{d-2}|\gamma_j|^{\alpha} - \left(|a_{1,d-1,d-1}|^{\alpha} + \sum_{j=0}^{d-2}|-\gamma_j|^{\alpha}\right) \\
 = & 0, 
\end{array}$$
where we used the independence and the unitary scale of the $Z_j$ along with Proposition~\ref{pro:sumSaS}.

The important conclusion of this section and of Theorem~\ref{thm:projMetric} is that the algorithm for predicting a future value of an LFSM given in Section~\ref{sec:condExp} also has some legitimacy when $\alpha\in(0,1)$. In what follows, we will thus use it to make some forecasts, whatever $\alpha\in(0,2]$.

\subsection{Quality of the forecast}\label{sec:quality}

Whatever the model used to forecast a financial time series, a traditional way of evaluating its quality is either by a root-mean-square error (RMSE), that is an $L^2$ norm, or by a hit ratio, that is the proportion of predictions in the good direction, provided we're interested in the binary problem of forecasting only if the time series is about to go up or down.

For the fBm, we have a theoretical expression both for the RMSE and the hit ratio~\cite{NP,Garcin2022}. When considering the LFSM instead of the fBm, some challenges appear. First, the pdf, which is required for calculating the hit ratio, can only be obtained by numerical means, namely by Fourier transform. Alternatively, one can also determine the hit ratio by simulations, as we do in Section~\ref{sec:simul}. 

The second challenge is about the RMSE, which is not defined for the LFSM when $\alpha<2$. But it is still possible to use a close metric with the $L^p$ norm of the residual, where $p<\alpha$. The Mellin transform is the central tool which makes it possible to have an explicit expression of the $L^p$ norm. Let's focus first on the case $\alpha=2$. The Mellin transform of the Gaussian distribution, using the change of variable $y=x^2/2$, is:
$$\begin{array}{ccl}
\int_0^{+\infty} x^{s-1}\frac{e^{-x^2/2}}{\sqrt{2\pi}}dx & = & \int_0^{+\infty} (x^2)^{(s-2)/2}\frac{e^{-x^2/2}}{\sqrt{2\pi}}xdx \\
 & = & \int_0^{+\infty} (2y)^{(s-2)/2}\frac{e^{-y}}{\sqrt{2\pi}}dy \\
 & = & \frac{2^{(s-2)/2}}{\sqrt{2\pi}}\Gamma\left(\frac{s}{2}\right).
\end{array}$$
From this, we easily get the $p$-th absolute moment of $X\sim\mathcal N(0,1)$, when $p>-1$:
$$\E\left(|X|^p\right)=2\int_0^{+\infty} x^{p}\frac{e^{-x^2/2}}{\sqrt{2\pi}}dx=\frac{2^{p/2}}{\sqrt{\pi}}\Gamma\left(\frac{p+1}{2}\right).$$
When $\alpha\in(0,2)$, we also know the absolute moment of $X$, an $S\alpha S$ variable of scale parameter 1, when $p\in(-1,\alpha)$, even though this result is not as straightforward and also requires the Fourier transform of the characteristic function~\cite{Hardin,Nolan2018,Shergin}:
$$\E\left(|X|^p\right)=\frac{\Gamma\left(1-\frac{p}{\alpha}\right)}{\Gamma(1-p)}\frac{1}{\cos\left(\frac{p\pi}{2}\right)},$$
noting that $\lim_{p\rightarrow 1}\Gamma(1-p)\cos(p\pi/2)=\pi/2$. As a consequence, if $(\mathcal T X_1,...,\mathcal T X_d)'$ is a vector of discrete-time observations of an $S\alpha S$ process, like an LFSM, admitting the decomposition provided in equation~\eqref{eq:transfodiscr}, if $\widehat{\mathcal T X}_d$ is the predictor of $\mathcal T X_d$ based on the values of $\mathcal T X_1$, ..., $\mathcal T X_{d-1}$, as defined in Section~\ref{sec:Predictor}, the residual of the forecast has the following $L^p$ norm:
$$\left(\E\left(\left|\mathcal T X_d-\widehat{\mathcal T X}_d\right|^p\right)\right)^{1/p} = |a_{1,d-1,d-1}|\left(\E\left(\left|Z_{d-1}\right|^p\right)\right)^{1/p} = |a_{1,d-1,d-1}|\left(\frac{\Gamma\left(1-\frac{p}{\alpha}\right)}{\Gamma(1-p)}\frac{1}{\cos\left(\frac{p\pi}{2}\right)}\right)^{1/p}.$$

\begin{figure}[htbp]
	\centering
		\includegraphics[width=0.45\textwidth]{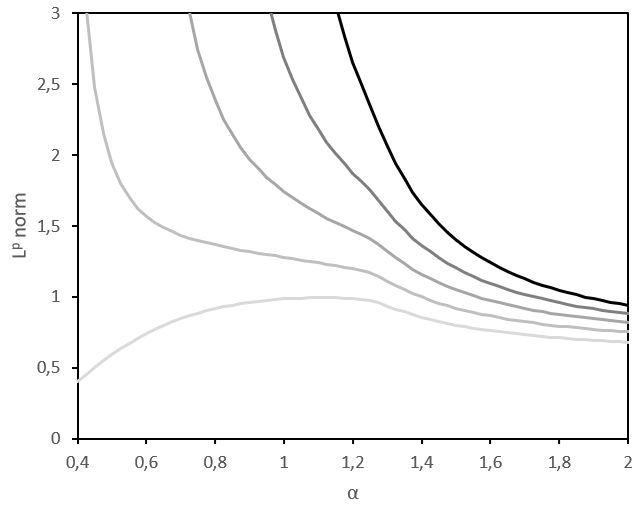} 
		\includegraphics[width=0.45\textwidth]{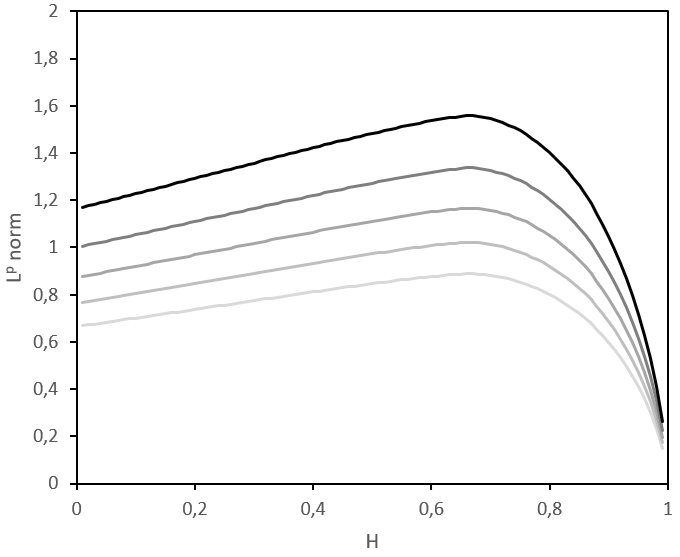} \\
		\includegraphics[width=0.45\textwidth]{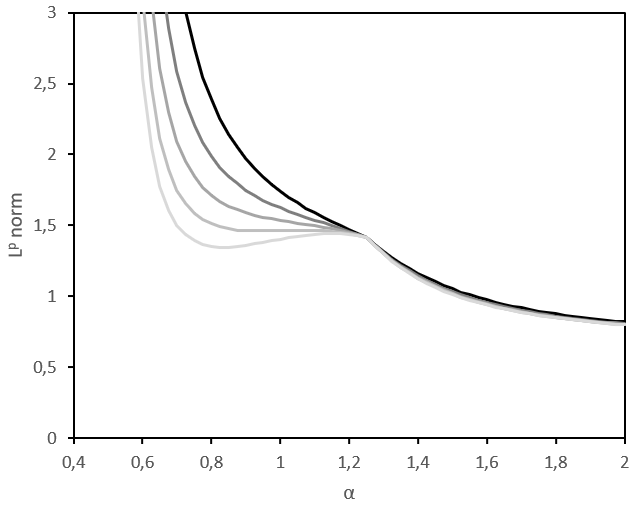} 
		\includegraphics[width=0.45\textwidth]{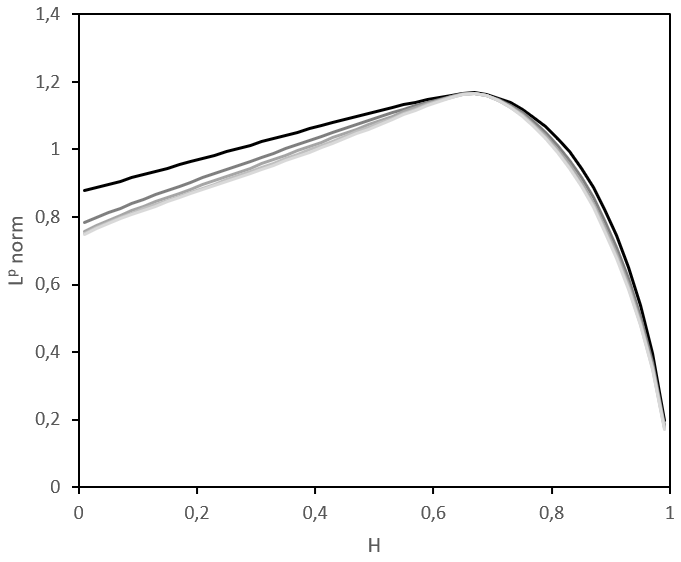} 
\begin{minipage}{0.7\textwidth}\caption{$L^p$ norm of the residual of the predictor $\widehat{\mathcal T X}_d$, either for various values of $\alpha\in(\max(0.4,p),2]$ and fixed $H=0.8$ (left graphs), or for various values of $H\in(0,1)$ and fixed $\alpha=1.5$ (right graphs). Each curve corresponds, from darkest to lightest, either to $p\in\{0.1, 0.3, 0.5, 0.7, 0.9\}$ and fixed $d=2$ (top graphs), or to $d\in\{2,4,9,16,32\}$ and fixed $p=0.5$ (bottom graphs).}
	\label{fig:ErreurTheorique}
\end{minipage}
\end{figure}

Figure~\ref{fig:ErreurTheorique} represents this $L^p$ norm error of the predictor $\widehat{\mathcal T X}_d$ for various values of $p$, $d$, $\alpha$, and $H$. We only consider values of $\alpha$ higher than $p$, leading to a finite $L^p$ norm. When fixing $H=0.8$, we also restrict the values of $\alpha$ so that we have a unique decomposition. Indeed, after Figure~\ref{fig:coefDecompo}, 0.4 is approximately the limit value for $\alpha$ that guarantees the uniqueness of the decomposition when $H=0.8$. Considering the curve of the error as a function of $\alpha$, a singularity appears for $\alpha=1/H$. Now considering the error as a function of $H$, a maximum is reached when $H=1/\alpha$. This is true only for $\alpha> 1$, otherwise the error is maximal at the limit $H\rightarrow 1$. Last, adding several observations to build the predictor, by (strongly) increasing $d$, even though it decreases the error, has a limited effect.

\section{Simulation study}\label{sec:simul}

\begin{figure}[htbp]
	\centering
		\includegraphics[width=0.45\textwidth]{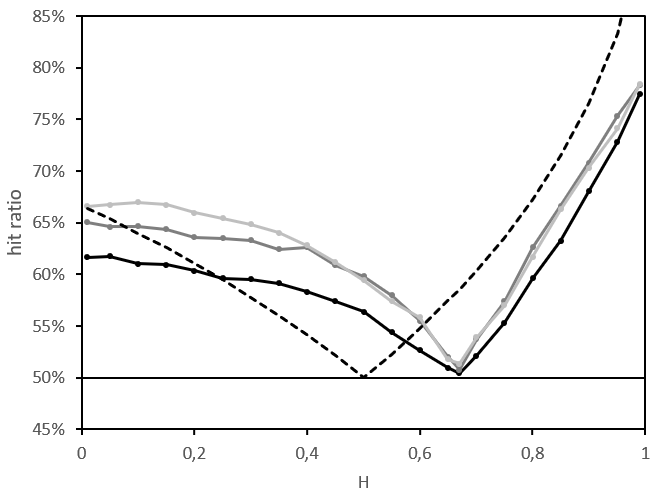} 
		\includegraphics[width=0.45\textwidth]{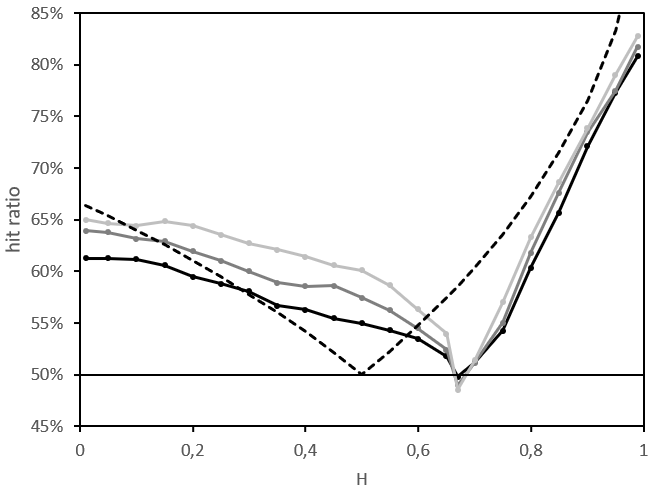} \\
		\includegraphics[width=0.45\textwidth]{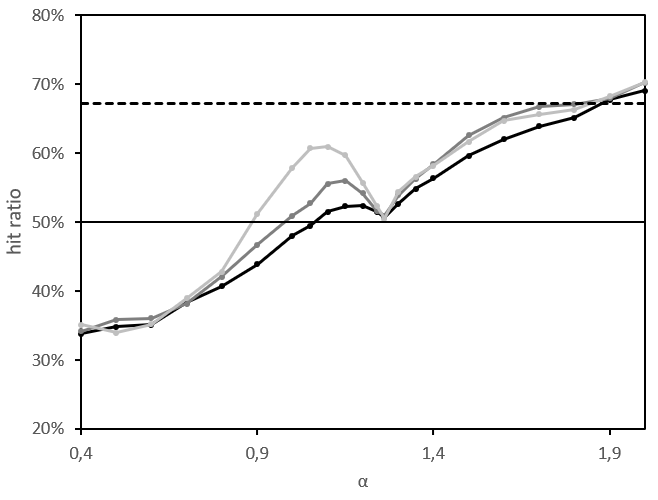} 
		\includegraphics[width=0.45\textwidth]{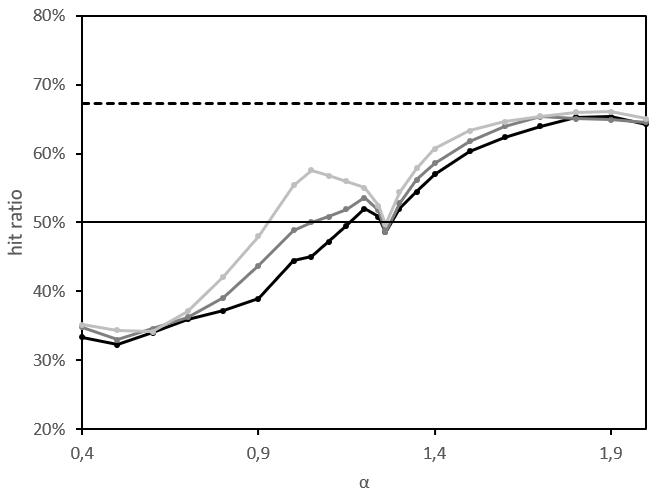} 
\begin{minipage}{0.7\textwidth}\caption{Hit ratio of the LFSM for $\alpha=1.5$ and $H\in(0,1)$ (top) or $\alpha\in[0.4,2]$ and $H=0.8$ (bottom), obtained by simulation. Each trajectory of LFSM, for a given pair $(\alpha,H)$, is simulated with a unique seed for the left graphs and a unique seed for the right graphs. The three curves correspond to $d=2$ (black), $d=5$ (dark grey), $d=20$ (light grey). The dotted curve is the theoretical hit ratio of an fBm with the same Hurst exponent and $d=2$, as expressed in equation~\eqref{eq:Hit_fBm}.}
	\label{fig:HitSimul}
\end{minipage}
\end{figure}

Since the hit ratio of the LFSM can only be obtained numerically, we conduct simulations to calculate it for various pairs $(\alpha,H)$, as displayed in Figure~\ref{fig:HitSimul}. For each pair of parameters, the hit ratio is calculated on a single, but long, simulated trajectory. We consider time series of length 2,001 and $d\in\{2,5,20\}$. Therefore, we have respectively 1,999, 1,995, and 1,981 forecasts for each time series. In order to get a smooth curve for the hit ratio when represented as a function of the parameters, we use the same seed to generate the pseudo-random numbers for each trajectory. We repeat the experiment with another seed and get very close graphs, as one can see in the right part of Figure~\ref{fig:HitSimul}, thus confirming the results. We also want to compare the obtained hit ratio to the theoretical one of an fBm with $d=2$, which is the $\rho$ defined by
\begin{equation}\label{eq:Hit_fBm}
\rho=1-\frac{1}{\pi}\arctan\left(\sqrt{\frac{1}{\left(2^{2H-1}-1\right)^2}-1}\right),
\end{equation}
for a given $H$~\cite{Garcin2022}.

For a fixed value of $\alpha=1.5$, the hit ratio of the LFSM, seen as a function of $H$, seems to be a translation to the right of the one of the fBm: the maximum is obtained for values of $H$ close to 1, depicting a strong and positive serial dependence, it is also higher than $50\%$ for $H$ closer to 0, meaning that a negative serial dependence can be exploited to forecast increments, and it reaches a minimum at $50\%$ for a value of $H$ that is $1/2$ for the fBm and $1/\alpha$ for the LFSM. This result illustrates that the sole value of $H$ is not enough to conclude about one's ability to forecast a time series: despite a value of $H$ equal to $1/2$, we have a hit ratio above $55\%$ when $\alpha=1.5$. Increasing (respectively decreasing) the value of $\alpha$ simply moves the minimum value of the curve to the left (resp. to the right).

We are now interested in the bottom graphs of Figure~\ref{fig:HitSimul}, where we consider a fixed value of $H=0.8$ and $\alpha$ in the range $[0.4,2]$. The restriction to this interval is because it guarantees that the decomposition exposed in Section~\ref{sec:decompo} is valid. When $\alpha$ is close to 2, we get a high hit ratio, close to the one of an fBm. Then, the hit ratio progressively decreases to $50\%$ when $\alpha$ decreases to $1/H$, value at which the curve reaches a local minimum. The shape of the curve for $\alpha<1/H$ is much more surprising: when $\alpha$ decreases below $1/H$, the hit ratio first increases above $50\%$, it reaches a local maximum, then it decreases below $50\%$. 

The hit ratio largely below $50\%$ that we obtain for very low values of $\alpha$ can be seen as a paradox, but we are able to explain it. For an fBm with $H<1/2$, non-overlapping increments are negatively correlated. Considering three consecutive increments, if the first one is positive, the second one, as negatively correlated to the first one, is more likely to be negative. The third one is thus negatively correlated to a positive and to a negative increment, but the correlation decreases rapidly in this framework where there is no long-range dependence. Therefore, the third increment will more likely be positive. In the case of an LFSM with $\alpha$ below $1/H$, the alternation of positive and negative increments is also very likely. But when, in addition, $\alpha$ is very small, the frequent occurrence of very large increments can disrupt the mechanism described above. Indeed, when one observes a very large increment, say a positive one, the next increment will more likely be a large, but not as large, negative number. Then, the third increment should be positive due to its negative codifference with the second one. But the first increment is so large that the negative codifference between the first and third increments will dominate the codifference between the second and third ones. Consequently, the third increment will more likely be negative. So the second and third increments give the illusion of being positively dependent, what we can explain, in causality terms, by the presence of an overwhelming confounder, namely the first and large increment. At the macroscopic scale, after a large positive increment, subsequent increments will be negative and their magnitude will gradually decrease. When considering the process itself instead of the increments, one thus observes after each large jump a kind of local trend or progressive recovery. This phenomenon is visible in the simulation displayed in the bottom left graph of Figure~\ref{fig:simul}. 

In the literature, $\alpha<1/H$ is associated to an absence of long-rang dependence~\cite[Section 7.4]{SamTaq}. But the explanation above shows that large increments, which occur for small values of $\alpha$, may disrupt this interpretation. This justifies why a forecast based on a limited number of past observations, thus neglecting some past large increments, performs poorly when $\alpha$ is very small.

Last, when one increases $d$ and thus enlarges the information set, Figure~\ref{fig:HitSimul} shows an improved performance, especially when $H<1/\alpha$, that is for a negative dependence of increments.

The above simulation study is based on an oracular framework, in which we use exactly the same parameters to simulate the dataset and to build the LFSM-based forecast. We now present a second and more realistic experiment, in which we ignore the parameters of the generating process. We thus estimate a model and make a forecast based on this estimate. 

The generating process is an LFSM of parameters $(\alpha,H)$ successively equal to $(1.5,0.3)$, $(1.5,0.5)$, $(1.5,0.8)$, $(1.8,0.3)$, $(1.8,0.5)$, and $(1.8,0.8)$. For each pair of parameters, we simulate a single trajectory of size 5,000. Then we estimate dynamically three models, in a rolling window of size 500: an LFSM, an fBm, and an autoregressive (AR) model on the increments of the simulated process with $d-1$ lags. The estimation of the AR model is based on a linear regression. The size of each trajectory and of the estimation sample is close to what is done in the empirical study of real volatility data, Section~\ref{sec:empirical}. For a given model, we use each estimate and $d$ lagged observations to forecast the observation right after the end of the estimation window, so that, for each trajectory, we have 4,500 out-of-sample point predictions. The accuracy of each model is assessed through the hit ratio and the mean absolute error (MAE) between the forecast and the true value.

\begin{figure}[htbp]
	\centering
		\includegraphics[width=0.45\textwidth]{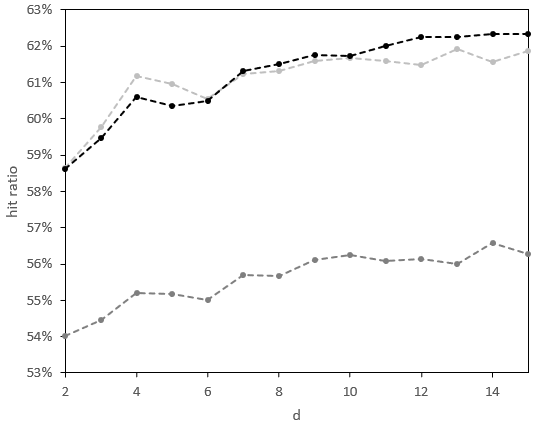} 
		\includegraphics[width=0.45\textwidth]{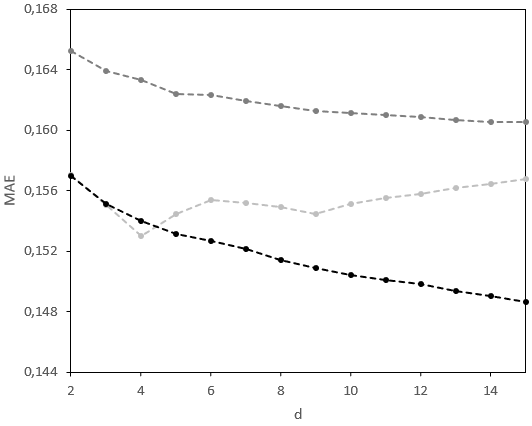} \\
		\includegraphics[width=0.45\textwidth]{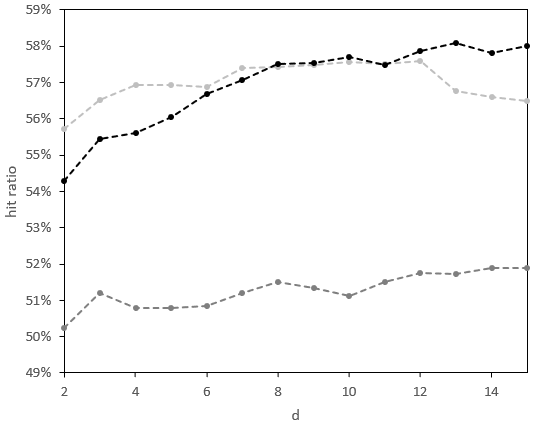} 
		\includegraphics[width=0.45\textwidth]{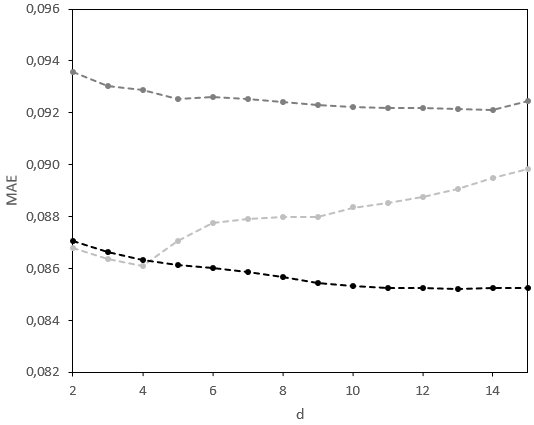} \\
		\includegraphics[width=0.45\textwidth]{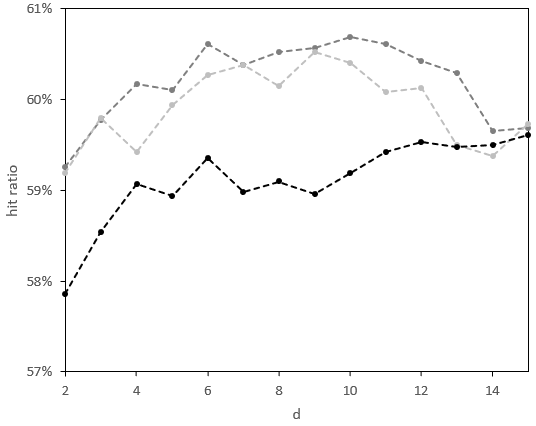} 
		\includegraphics[width=0.45\textwidth]{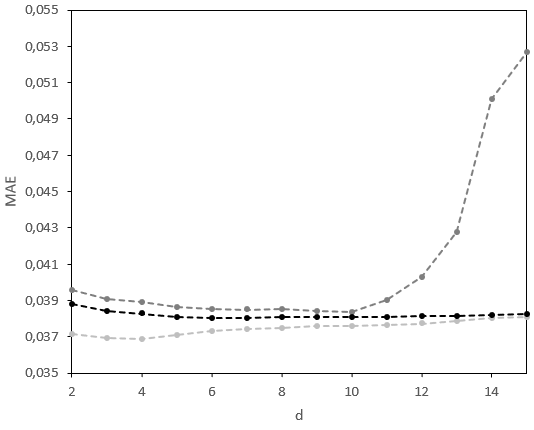} \\
\begin{minipage}{0.7\textwidth}\caption{Hit ratio (left) and MAE (right) for a trajectory of 5,000 observations generated by an LFSM of parameters $\alpha=1.5$ and $H$ equal to 0.3 (top graphs), 0.5 (middle graphs), and 0.8 (bottom graphs). The series is forecast either with an LFSM (black), or with an fBm (dark grey), or with an AR model (light grey), for $d\in\llbracket 2,15\rrbracket$ after estimation of these models in a rolling window of 500 observations.}
	\label{fig:SimulEstim_alpha15}
\end{minipage}
\end{figure}

Figure~\ref{fig:SimulEstim_alpha15} gathers the results for $\alpha=1.5$. The three models display a satisfying hit ratio, above $50\%$. When $H\in\{0.3,0.5\}$, case for which the autocodifference is negative, the fBm is the worst of the three models whereas the LFSM and the AR model are close to each other, at least regarding the hit ratio. When $d$ increases, the LFSM outperforms the AR model, both regarding the hit ratio criterion, from $d\geq 12$, and more strongly regarding the MAE, from $d\geq 5$. The poor MAE of the AR model when $d$ is large comes from the overfitting of this model, since the number of its parameters increases with $d$, whereas both the fBm and the LFSM have the same number of parameters whatever $d$. When $H=0.8$, the memory $H-1/\alpha$ is slightly above zero. Since it is close to zero, the uncertainty in the estimation of the parameters of the LFSM may locally lead to an estimate with a negative autocodifference and thus a less good forecast. For this reason, when $(\alpha,H)=(1.5,0.8)$, the LFSM slightly underperforms the two other models, even though, for large values of $d$, the LFSM is on par with the AR model and has a much lower MAE than the fBm.

\begin{figure}[htbp]
	\centering
		\includegraphics[width=0.45\textwidth]{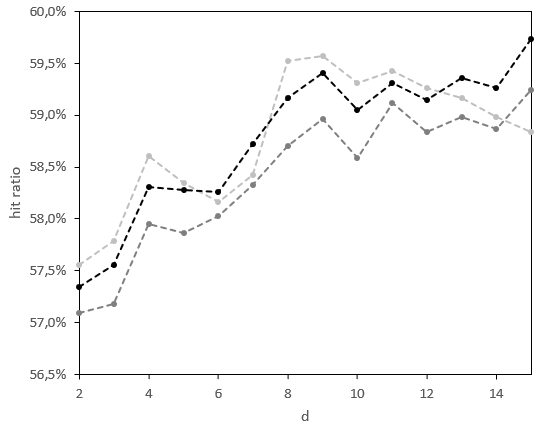} 
		\includegraphics[width=0.45\textwidth]{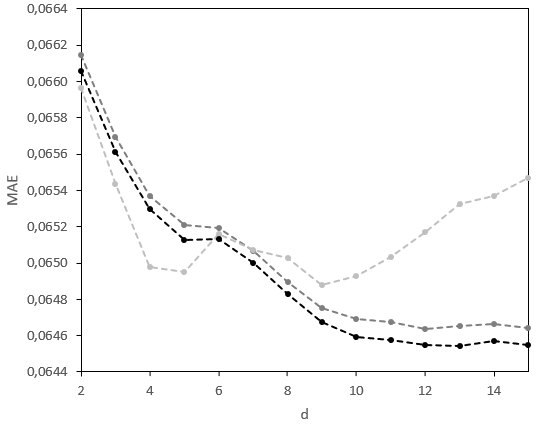} \\
		\includegraphics[width=0.45\textwidth]{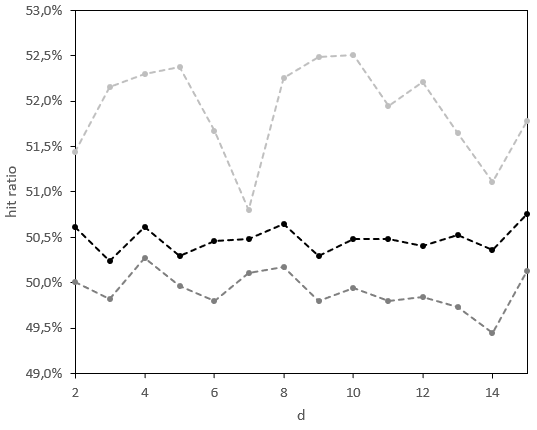} 
		\includegraphics[width=0.45\textwidth]{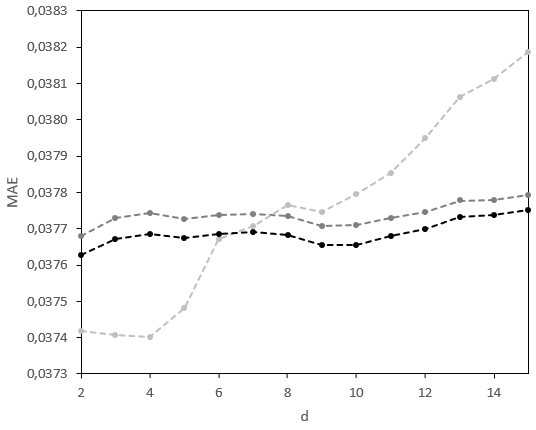} \\
		\includegraphics[width=0.45\textwidth]{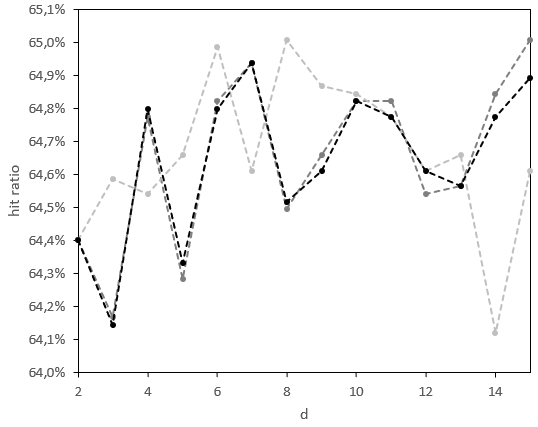} 
		\includegraphics[width=0.45\textwidth]{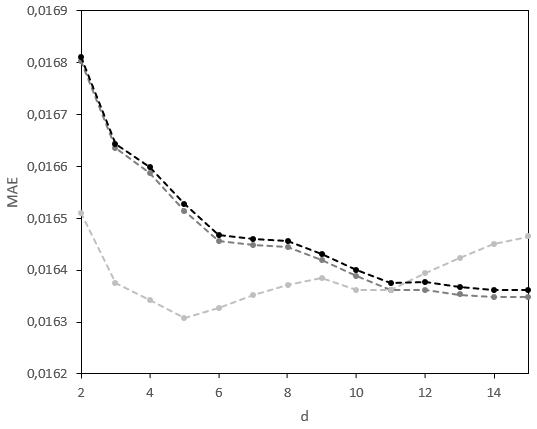} \\
\begin{minipage}{0.7\textwidth}\caption{Hit ratio (left) and MAE (right) for a trajectory of 5,000 observations generated by an LFSM of parameters $\alpha=1.8$ and $H$ equal to 0.3 (top graphs), 0.5 (middle graphs), and 0.8 (bottom graphs). The series is forecast either with an LFSM (black), or with an fBm (dark grey), or with an AR model (light grey), for $d\in\llbracket 2,15\rrbracket$ after estimation of these models in a rolling window of 500 observations.}
	\label{fig:SimulEstim_alpha18}
\end{minipage}
\end{figure}

When $\alpha=1.8$, we see in Figure~\ref{fig:SimulEstim_alpha18} the same behaviour as the one described above: overall, the LFSM is close to both the fBm and the AR model and it outperforms these two models when $d$ is large. The slightly less favourable case for LFSM is when $H=0.5$, which is the one for which the memory is very slightly positive, like for $(\alpha,H)=(1.5,0.8)$. It indeed again leads to an estimate with uncertain autocodifference. However, when $d$ increases, the MAE criterion indicates that the LFSM is the best of the three models. The parameter set $(\alpha,H)=(1.8,0.5)$ is also the only one in our experiment leading to a hit ratio sometimes lower than $50\%$ and only for the fBm. Indeed, whatever $\alpha$, if $H=0.5$ the estimated fBms have a very fluctuating sign of autocovariance and the resulting predictions are highly uncertain.

\section{Empirical application to rough volatility}\label{sec:empirical}

In this section, we investigate the performance of the forecasting method based on LFSM when applied to real data. We focus on financial data, namely daily time series of volatilities. We make a special effort to give an empirical evidence of the relevance of this non-Gaussian fractional model in this framework.

As explained at the end of Section~\ref{sec:estim}, in real applications, we use the model $Y_t=\sigma X_t$, where $X_t$ is an LFSM. But our forecasting method is based on coefficients $a_{t,i,j}$ that are determined for a standard LFSM. Therefore, we apply our method to the time series $Y_t/\widehat \sigma$, where $\widehat \sigma$ is the estimate of $\sigma$, obtained as explained in Section~\ref{sec:estim}.

Modelling log-volatility with an fBm is an old idea~\cite{CR} which has recently seen a resurgence of interest in the mathematical finance community, since the idea of using a Hurst exponent lower than $1/2$ emerged~\cite{ALV,GJR}, making it possible to depict rough trajectories of volatilities. Since then, the empirical relevance of the model has been studied~\cite{GG,CD,AJL} and extensions proposed, like the addition of jumps~\cite{AJC} or of a dependence on other volatilities~\cite{BYZ}. In the same time, forecasting methods based on these fractional models of volatility have been developed~\cite{Garcin2022,BYZ}.

We propose here to study the relevance of replacing the Gaussian distribution of the fBm by an $\alpha$-stable one and to apply the forecasting method of the LFSM exposed in Section~\ref{sec:Predictor} to time series of log-volatilities. The data used in our analysis are the same as those used in a previous work on forecasting volatilities with an fBm~\cite{Garcin2022}, namely daily realized volatilities computed with a five-minute discretization of prices, imported from the formerly available Oxford-Man Institute of Quantitative Finance Realized Library. We focus on the realized volatility of eight stock indices: the AEX index, the CAC 40 index, the FTSE 100 index, the Nasdaq 100 index (IXIC), the Nikkei 225 index (N225), the Oslo Exchange All-share index (OSEAX), the Madrid General index (SMSI), and the S\&P 500 index. The series starts on January 2000, except N225, which starts in February 2000, OSEAX in September 2001, and SMSI in July 2005. The end date of our sample is on the 12th April 2021. 

The papers that justify the empirical relevance of the fBm for depicting log-volatilities mainly focus on the fractal properties of the time series and often overlook the justification of the Gaussian distribution. Instead, Table~\ref{tab:signifVolGauss} shows that series of the increments of the logarithm of realized volatilities have a significant non-Gaussian behaviour, with a leptokurtic distribution. First, the Jarque-Bera test, which is based on both the skewness and the kurtosis, as well as a Z-test based on the sole kurtosis, both reject with a strong confidence the null hypothesis of Gaussian increments. Second, using the extreme value theory (EVT), we study the speed at which the tails of the density go to zero. For $N$ observations, it is based on those ranked $N-k_N$, $N-2k_N$, and $N-4k_N$, where the conditions $k_N\rightarrow\infty$ and $k_N/N\rightarrow 0$ ensure the consistency of Pickands estimator $\hat\xi_{k_N}$ of the EVT parameter. In practice, the estimator is very sensitive to the choice of the free parameter $k_N$, so that we define a more robust estimator $\hat\xi$ as the median of $\hat\xi_1$, ...,$\hat\xi_{40}$.\footnote{ In the application, $N$ is around $5,000$, with some variations depending on the series.} Simulating $10,000$ times $N$ Gaussian variables, we also get an approximate distribution of these estimators under the null hypothesis of Gaussian increments. Table~\ref{tab:signifVolGauss} reveals a distinct behaviour in the right and left tails: although on the left the null hypothesis can only be rejected for SPX, on the right the Fréchet distribution and therefore fat tails are validated for all the series except IXIC. While the observations show some asymmetry, we restrict attention here to the symmetric LFSM, focusing on the impact of heavy tails and serial dependence. Extensions to asymmetric models are left for future work.

\begin{table}[htbp]
\centering
\begin{tabular}{|l|c|c|c|c|c|c|c|}
\hline
series & JB statistic & kurtosis & Z-test statistic & EVT right & EVT left \\
\hline
 AEX & $1303^{\star\star\star}$ & 5.41 & $36.1^{\star\star\star}$ & $0.180^{\star}$ & -0.058\\
 CAC 40 & $1154^{\star\star\star}$ & 5.25 & $33.9^{\star\star\star}$ & $0.381^{\star\star}$ & 0.129 \\
 FTSE & $416^{\star\star\star}$ & 4.35 & $20.5^{\star\star\star}$ & $0.181^{\star}$ & -0.014 \\
 IXIC & $233^{\star\star\star}$ & 3.97 & $14.4^{\star\star\star}$ & -0.125 & -0.167 \\
 N225 & $1183^{\star\star\star}$ & 5.21 & $32.5^{\star\star\star}$ & $0.422^{\star\star\star}$ & 0.014 \\
 OSEAX & $295^{\star\star\star}$ & 4.19 & $17.0^{\star\star\star}$ & $0.179^{\star}$ & -0.275 \\
 SMSI & $1026^{\star\star\star}$ & 5.45 & $31.7^{\star\star\star}$ & $0.441^{\star\star\star}$ & 0.029 \\
 SPX & $150^{\star\star\star}$ & 3.80 & $11.8^{\star\star\star}$ & $0.478^{\star\star\star}$ & $0.233^{\star}$ \\
\hline
\end{tabular}
\begin{minipage}{0.7\textwidth}\caption{For each series of realized volatility, Jarque-Bera (JB) statistic, kurtosis along with the statistic of the corresponding Z-test, and EVT parameter obtained by Pickands' estimator, both for right and left tails. The significance at $95\%$ ($^{\star}$), $99\%$ ($^{\star\star}$), and $99.9\%$ ($^{\star\star\star}$) is indicated for the Jarque-Bera test, the Z-test, and the tests on the EVT parameters.}
\label{tab:signifVolGauss}
\end{minipage}
\end{table}

\begin{figure}[htbp]
	\centering
		\includegraphics[width=0.6\textwidth]{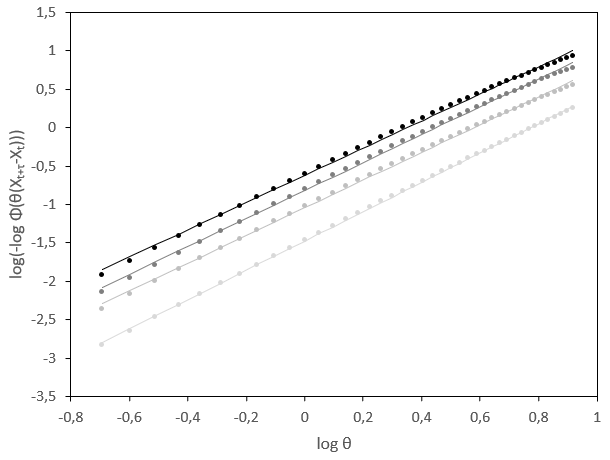} 
\begin{minipage}{0.7\textwidth}\caption{$\ln(-\ln(\widehat \Phi_{X_{.+\tau}-X_.}(\theta)))$ as a function of $\ln|\theta|$, along with its linear regression, for $X$ the full series of the logarithm of the realized volatility on SPX, $\theta\in[0.5,2.5]$, and, from bottom to top, with a duration of increment $\theta$ equal (in days) to 1, 5, 10, and 22. }
	\label{fig:SPX_foncChar}
\end{minipage}
\end{figure}

Once the Gaussian distribution is rejected, we can assess the relevance of the $\alpha$-stable alternative by giving some details on the estimation. Indeed, the estimator proposed in Section~\ref{sec:estim} is based on a linear regression of a specific transformation of the empirical characteristic function $\widehat \Phi_{X_{.+\tau}-X_.}$ of increments, for a reasonable set of parameters $\theta$ and a fixed duration $\tau$. We can see in Figure~\ref{fig:SPX_foncChar}, based on the SPX series, that the linear shape is satisfying, whatever the duration $\tau$, so that one cannot reject the $\alpha$-stable hypothesis. The slope directly provides us with an estimate of $\alpha$, equal to 1.91 in the case the most subject to noise, that is $\tau=1$, and equal to 1.80, 1.82, and 1.77 for $\tau\in\{5,10,22\}$.

The second purpose of this empirical study is to forecast the next daily variation of volatility. For each series, and each day $t$, we estimate the parameters of an LFSM in the two-year window finishing at $t$. Next, using these parameters and the method developed in Section~\ref{sec:Predictor}, we forecast the log-volatility of day $t+1$. We first compare the result with a forecast made by fBm, using the same estimation sample. 

\begin{figure}[htbp]
	\centering
		\includegraphics[width=0.45\textwidth]{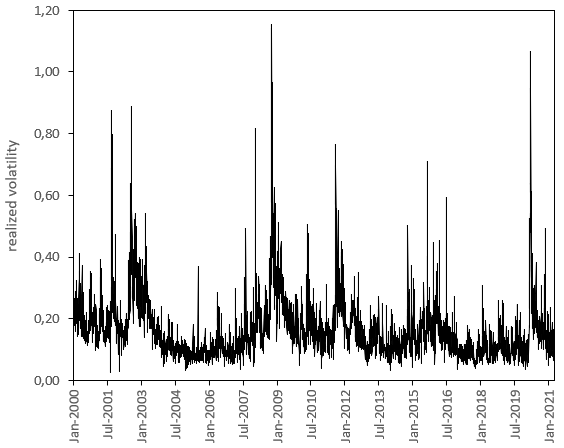} 
		\includegraphics[width=0.45\textwidth]{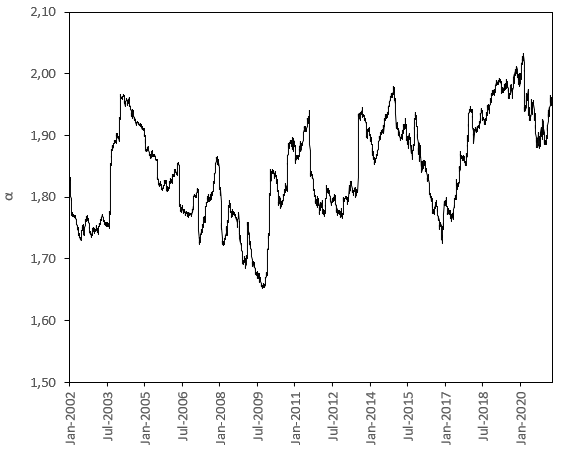} \\
		\includegraphics[width=0.45\textwidth]{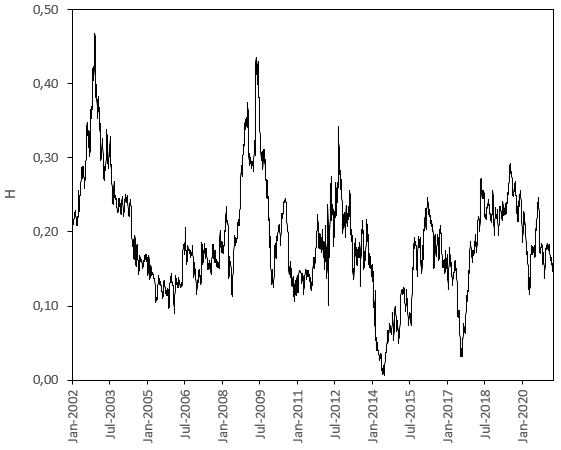} 
		\includegraphics[width=0.45\textwidth]{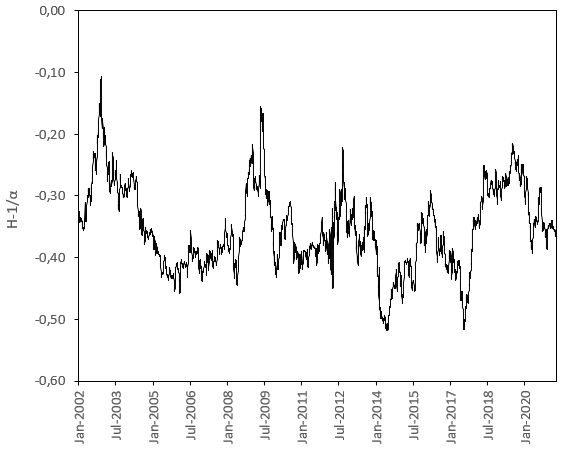} 
\begin{minipage}{0.7\textwidth}\caption{Time series of annualized realized volatility of the CAC 40 index (top left), estimated $\alpha$ (top right), $H$ (bottom left), and $H-1/\alpha$ (bottom right) of log-process, using a two-year rolling window.}
	\label{fig:CAC}
\end{minipage}
\end{figure}

Figure~\ref{fig:CAC} displays the trajectory of the realized volatility of the CAC 40 index, along with the dynamic estimates of the parameters $\alpha$ (between 1.65 and 2) and $H$ (lower than $1/2$) of the log-process. We also represent $H-1/\alpha$, that is the memory of the log-process, which, as one can see in Figure~\ref{fig:CAC}, is always negative. It means that the sign of the forecast of the future increment of volatility is simply the opposite of the one of the last volatility increment, if the forecast is based on this sole lagged observation. In other words, when $d=2$ the LFSM and the fBm will lead to the same hit ratios. 

\begin{figure}[htbp]
	\centering
		\includegraphics[width=0.45\textwidth]{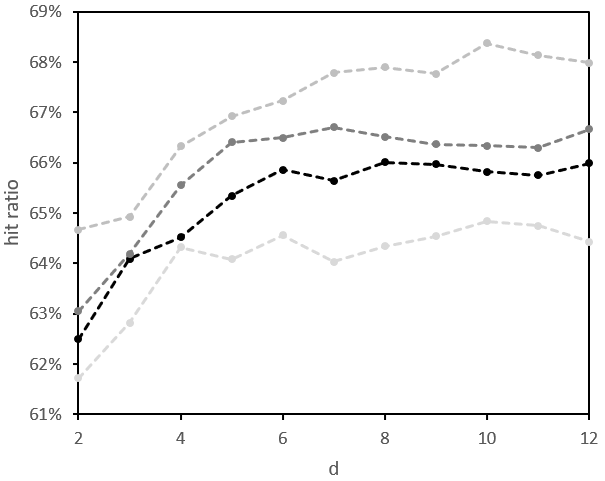} 
		\includegraphics[width=0.45\textwidth]{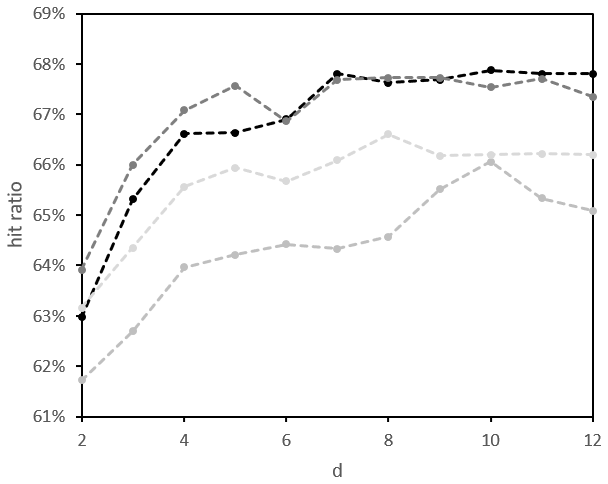}
\begin{minipage}{0.7\textwidth}\caption{Hit ratio of the forecast of the next daily variation of volatility, for $d\in\llbracket 2,12\rrbracket$. The curves correspond to the following indices, from the darkest to the lightest: AEX, CAC 40, FTSE, IXIC (left graph), and N225, OSEAX, SMSI, SPX (right graph).}
	\label{fig:vol_hit}
\end{minipage}
\end{figure}

Things may be different when one considers a larger information set, that is $d>2$. We gather the hit ratio for various values of $d$ in Figure~\ref{fig:vol_hit}. It shows a good ability of the LFSM to forecast future values of realized log-volatility, with hit ratios approximatively between $62\%$ and $68\%$, in general larger when $d$ increases. This result validates the method. However, we don't see a big difference with the performance of the forecasting method based on fBm~\cite{Garcin2022}: depending on the time series, the average, over all the parameters $d$, of the absolute difference between the hit ratios obtained in the LFSM and the fBm frameworks, is between $0.1\%$ et $0.3\%$. For some time series (IXIC and OSEAX), the fBm always performs slightly better than the LFSM; for others (N225 and SMSI), it is the opposite.

\begin{figure}[htbp]
	\centering
		\includegraphics[width=0.45\textwidth]{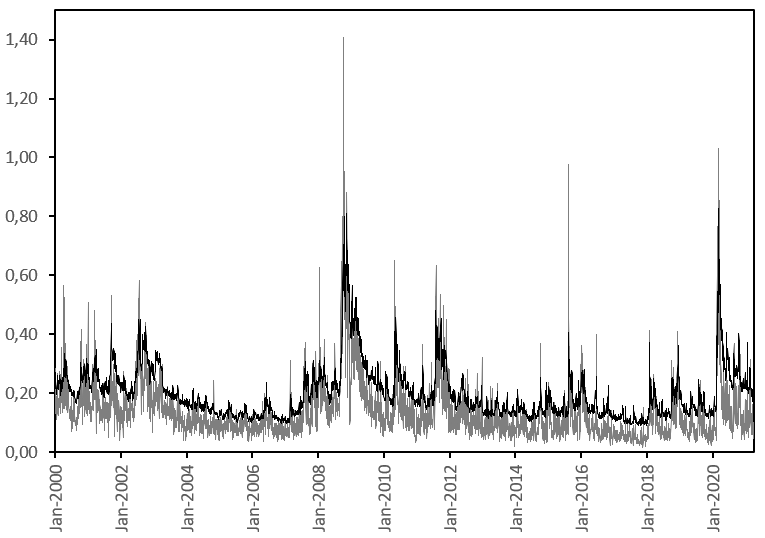} 
		\includegraphics[width=0.45\textwidth]{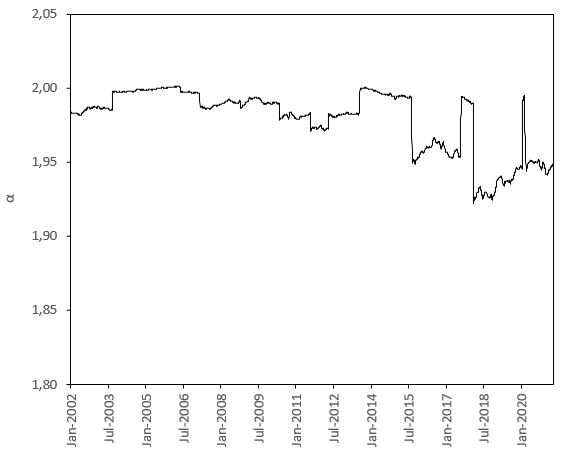} \\
		\includegraphics[width=0.45\textwidth]{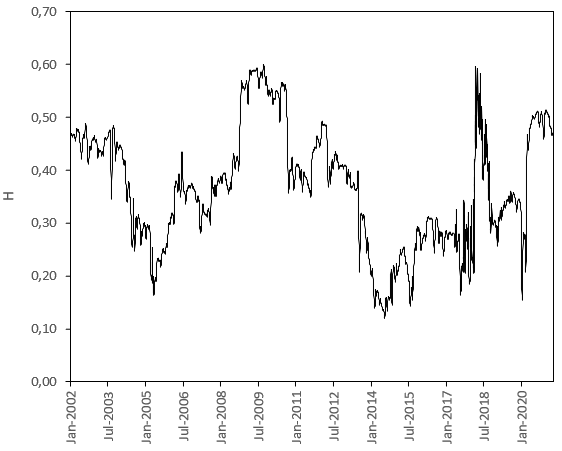} 
		\includegraphics[width=0.45\textwidth]{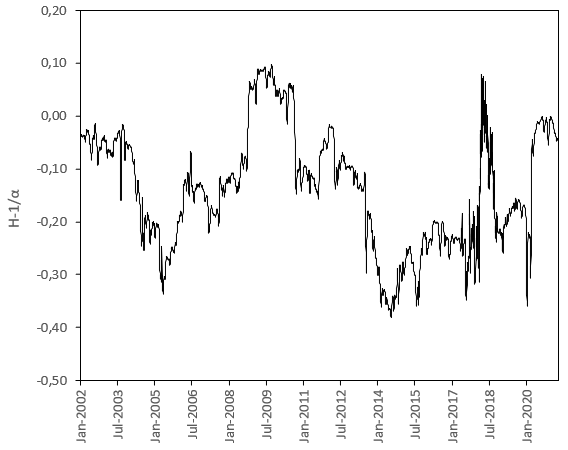} 
\begin{minipage}{0.7\textwidth}\caption{Time series of VIX (divided by 100) in black and of the annualized realized volatility of the SPX index in grey (top left), estimated $\alpha$ (top right), $H$ (bottom left), and $H-1/\alpha$ (bottom right) of the log-VIX time series, using a two-year rolling window.}
	\label{fig:VIX}
\end{minipage}
\end{figure}

Beyond realized volatilities, we are also interested in the VIX series. We import from Bloomberg the daily close price of VIX between January 2000 and 12th April 2021. Empirical studies on rough volatility rarely consider this series of volatility. The Hurst exponent of the log-VIX series is indeed not similar to the one estimated for realized log-volatilities, with a value much closer to $1/2$~\cite{ZC}. Figure~\ref{fig:VIX} indeed shows that the time series of VIX looks like a smoothed version of the realized volatility of SPX. The Hurst exponent is between 0.1 and 0.6 and oscillates around 0.4. The parameter $\alpha$ is only slightly lower than 2. However, extreme events are not as rare as in a Gaussian distribution: the Jarque-Bera statistic above 10,000 and the kurtosis at 9.78 reject the Gaussian hypothesis with a confidence larger than $99.9\%$. There is also an asymmetry in the tails, with an EVT parameter on the right significantly above 0 (0.163) and an EVT parameter on the left not significantly different from 0 (-0.030).

\begin{figure}[htbp]
	\centering
		\includegraphics[width=0.6\textwidth]{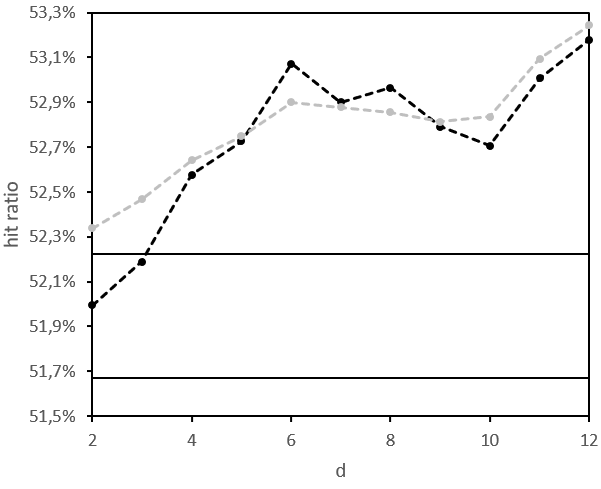} 
\begin{minipage}{0.7\textwidth}\caption{Hit ratio of the forecast of the next daily variation of VIX with an LFSM (black) or an fBm (grey), for $d\in\llbracket 2,12\rrbracket$. The two thin solid lines correspond to the confidence bound at $99\%$ and $99.9\%$ of a binomial test, for 4856 predictions.}
	\label{fig:VIX_hit}
\end{minipage}
\end{figure}

Since $H-1/\alpha$ is much closer to zero than in the rough volatility paradigm, forecasts are less accurate, consistently with the observations made in Section~\ref{sec:simul}. Figure~\ref{fig:VIX_hit} shows that the fBm is outperforming when $d$ is 2 or 3, whereas the LFSM makes better predictions with $d$ equal to 6 or 8. Overall, the hit ratio is significantly above $1/2$ when $d$ is large enough, namely $d\geq 4$.

To complement this study, we want to compare the forecasts made by LFSM and fBm with a widespread alternative, namely the HAR model. By combining observations at daily, weekly, and monthly scales, HAR partly captures daily variations and long-range dependence~\cite{Corsi,CAR}. It is considered as one of the most performing forecasting volatility models. The HAR forecast writes
$$\widehat X_{t+1}^\text{HAR}=\beta_0+\beta_1 X_t+ \beta_2 \sum_{i=0}^4 X_{t-i}+ \beta_3 \sum_{i=0}^{21} X_{t-i},$$
where $X_t$ is the logarithm of the realized volatility defined at the five-minute sampling, $\beta_0$, $\beta_1$, $\beta_2$, and $\beta_3$ are parameters estimated through a linear regression in a two-year window finishing at time $t$. To make a fair comparison, we also select $d=22$ as the number of lags in the fBm and LFSM forecasts.

Table~\ref{tab:signifVol_MAE} gathers the MAE for the forecast made by our three models based each on 22 past observations. Since the three models considered are all distributed-lag models with the same number of lags, their difference only relies on the weighting function used: either piecewise-constant for the HAR model, or based on power functions for the fBm and the LFSM. This naturally leads to close forecasting performances. However, a clear ranking of the three models appears, according to the MAE criterion: the HAR model always has the lowest MAE, followed by the LFSM and by the fBm. We assess the significance of the difference in predictive accuracy using a Diebold-Mariano and West (DMW) test~\cite{DM,West}, based on the MAE loss. This loss function is well suited to our leptokurtic setting, as it limits the influence of large volatility spikes on the evaluation. Our DMW test compares either the HAR or the fBm with the LFSM. The statistic is negative when the LFSM performs worse and positive when it performs better. Under the null hypothesis of equal predictive accuracy, the test statistic is asymptotically standard normal. We find that the HAR model outperforms the LFSM, and the LFSM outperforms the fBm, but these differences are statistically significant for only 4 out the 9 series.

\begin{table}[htbp]
\centering
\begin{tabular}{|l|c|c|c|}
\hline
series & HAR & fBm & LFSM \\
\hline
AEX & 0.393 (-4.26$^{\star\star\star}$) & 0.399 (1.91) & 0.397 \\
CAC 40 & 0.390 (-3.54$^{\star\star\star}$) & 0.396 (2.32$^{\star}$) & 0.394 \\
FTSE & 0.448 (-1.40) & 0.452 (2.96$^{\star\star}$) & 0.449 \\
IXIC & 0.414 (-1.78) & 0.424 (0.91) & 0.423 \\
N225 & 0.413 (-3.73$^{\star\star\star}$) & 0.416 (1.97$^{\star}$) & 0.415 \\
OSEAX & 0.456 (-0.23) & 0.458 (2.36$^{\star}$) & 0.457 \\
SMSI & 0.388 (-0.90) & 0.390 (0.70) & 0.390 \\
SPX & 0.472 (-2.34$^{\star}$) & 0.482 (0.84) & 0.481 \\
VIX & 0.050 (-1.67) & 0.052 (-1.42) & 0.052 \\
\hline
\end{tabular}
\begin{minipage}{0.7\textwidth}\caption{For each series of realized log-volatility, as well as for the log-VIX, MAE and, in parenthesis, DMW statistic against the LFSM for the MAE loss. The significance at $95\%$ ($^{\star}$), $99\%$ ($^{\star\star}$), and $99.9\%$ ($^{\star\star\star}$) is indicated for the DMW test.}
\label{tab:signifVol_MAE}
\end{minipage}
\end{table}

When it comes to hit ratios, LFSM performs better than fBm for 6 series, but the difference is never significant, according to a McNemar test. Comparing LFSM and HAR, HAR performs significantly better for CAC 40 and SPX, whereas LFSM performs better for FTSE, as one can see in Table~\ref{tab:signifVol_hit}. Other series do not show a significant difference of performance accuracy between HAR and LFSM.

\begin{table}[htbp]
\centering
\begin{tabular}{|l|c|c|c|}
\hline
series & HAR & fBm & LFSM \\
\hline
AEX & 66.43$\%$ (0.655) & 66.23$\%$ (1.000) & 66.23$\%$ \\
CAC 40 & 67.70$\%$ (0.055) & 66.69$\%$ (0.419) & 66.83$\%$ \\
FTSE & 67.39$\%$ (0.023) & 68.38$\%$ (0.895) & 68.40$\%$ \\
IXIC & 64.97$\%$ (0.224) & 64.29$\%$ (0.583) & 64.39$\%$ \\
N225 & 67.31$\%$ (0.458) & 67.48$\%$ (0.310) & 67.65$\%$ \\
OSEAX & 68.07$\%$ (0.716) & 68.39$\%$ (0.406) & 68.23$\%$ \\
SMSI & 65.51$\%$ (0.525) & 65.68$\%$ (0.307) & 65.88$\%$ \\
SPX & 67.51$\%$ (0.011) & 66.27$\%$ (0.686) & 66.33$\%$ \\
VIX & 53.83$\%$ (0.840) & 53.81$\%$ (1.000) & 53.67$\%$ \\
\hline
\end{tabular}
\begin{minipage}{0.7\textwidth}\caption{For each series of realized log-volatility, as well as for the log-VIX, hit ratio and, in parenthesis, p-value of a McNemar test against the LFSM.}
\label{tab:signifVol_hit}
\end{minipage}
\end{table}

\begin{figure}[htbp]
	\centering
		\includegraphics[width=0.45\textwidth]{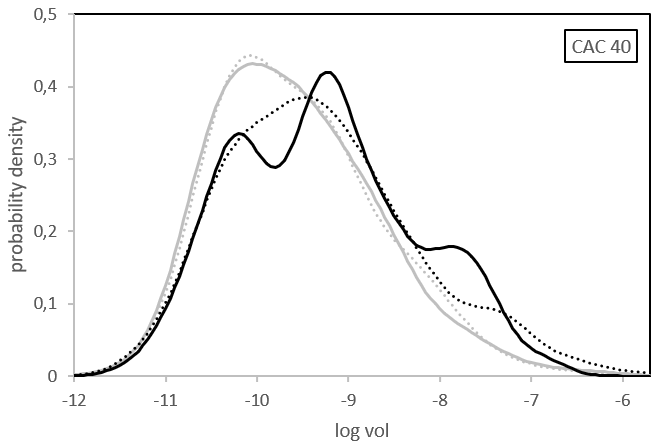} 
		\includegraphics[width=0.45\textwidth]{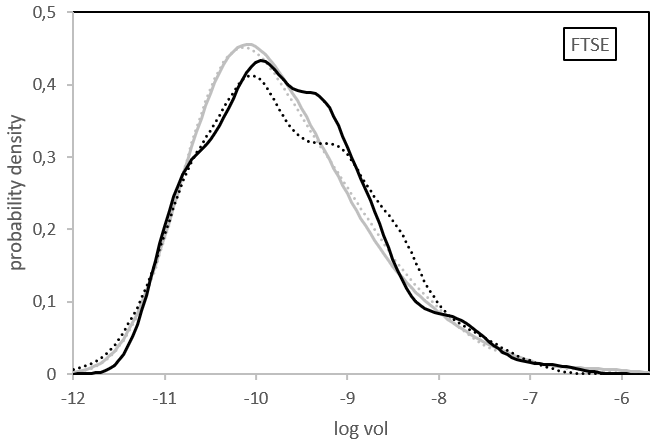} \\
		\includegraphics[width=0.45\textwidth]{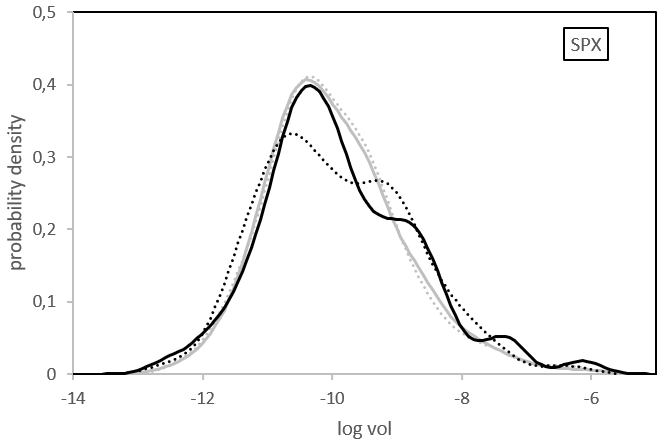} 
		\includegraphics[width=0.45\textwidth]{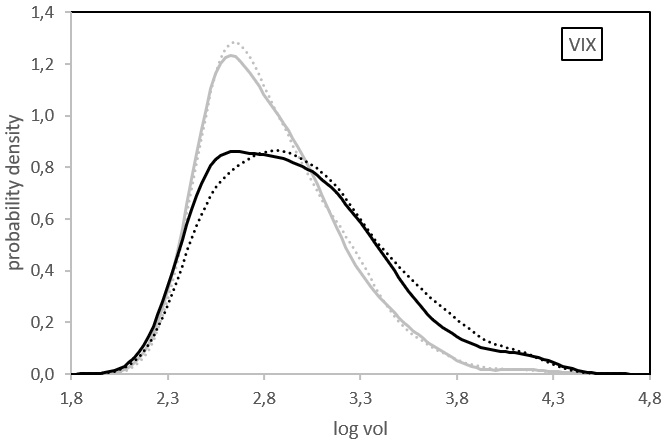}  
\begin{minipage}{0.7\textwidth}\caption{Kernel density estimation of the probability density of the five-day average of the logarithm of the realized volatility (or of the log-VIX, bottom right), conditionally on the quality of the forecast. Four categories are considered for the conditioning: when LFSM is right and HAR wrong (black solid line), the reverse (black dotted line), when both are right (grey solid line), or when both are wrong (grey dotted line). The stock indices considered are CAC 40 (top left), FTSE (top right), SPX (bottom left). The bandwidth used for each curve is determined by the maximum-complexity method~\cite{Garcin2024}.}
	\label{fig:KDE}
\end{minipage}
\end{figure}

Taken together, these results indicate that the three models have broadly comparable predictive accuracy, with variations across time series. Nevertheless, a general ranking emerges, with the HAR model tending to outperform the LFSM, and the LFSM the fBm. In order to more deeply analyze the difference between HAR and LFSM, we classify the joint outcome: a category when both models are right, an other when both are wrong, and two categories corresponding to a divergence between the two forecasts. We then try to link this outcome to the market regime that prevailed at the time of the forecast, which is characterized by a smooth version of the current log-volatility, namely the average of the last five daily observations. We see in Figure~\ref{fig:KDE} the kernel density of this recent log-volatility conditional on the four outcomes. For example, the peak of the black curve when the log-volatility of the CAC 40 is around -7.8, indicates that LFSM performs relatively well under this high-volatility regime. Whatever the series, we don't observe a big difference between the densities when both models agree (grey curves). The divergence between the models (black curves) more frequently occurs for larger volatilities, with a more pronounced effect for the CAC 40 and the VIX. Combining these densities and the number of observations in each category, we are able to build ranges of log-volatilites for which the LFSM performs better than the HAR model, even for series for which HAR is in average more accurate than LFSM. To assess the significance of this conditional outperformance, we conduct again a McNemar test limited to the observations in the range. For example, we find that when we observe a past log-vol of the CAC 40 in the range $[-8,-7.35]$, that is an annualized vol in $[29.5\%,40.9\%]$ (there are 41 such observations that lead to divergent forecasts), the LFSM performs significantly better than the HAR model (p-value of 0.019). Similarly, when the average five past observations of the logarithm of the VIX is in the range $[1.8,2.65]$, that is for an average VIX in $[6,14.2]$ (401 such observations leading to divergent forecasts), the LFSM also performs significantly better than the HAR model for the VIX forecast (p-value of 0.041). These results indicate that LFSM and HAR are complementary models and so that LFSM can improve volatility forecasts based only on the HAR model. 

\section{Conclusion}\label{sec:conclusion}

We have seen that the traditional method for forecasting an fBm, based on the covariance matrix, is not relevant when the considered process is $\alpha$-stable, like the LFSM. Instead, the codifference  can be used as a measure of serial dependence, even though it does not capture the entire dependence structure, which is more thoroughly described by the spectral measure. We have proposed a way of decomposing discrete-time observations of an LFSM in a sum of independent $\alpha$-stable variables. More precisely, the resulting decomposition has the same codifference as the LFSM but not the same spectral measure. We have shown that, under some conditions on the parameters $\alpha$ and $H$, this decomposition is unique. It can be used to propose a forecast of a future increment of the LFSM, defined either as a conditional expectation if $\alpha>1$, or as a semimetric projection otherwise. We have also been able to quantify the accuracy of the method, either theoretically with the $L^p$ norm of the error, or numerically with the hit ratio. Extending to the LFSM the interpretation of the Hurst exponent in the Gaussian case, we have been able to identify four regimes, instead of three: persistence of the increments when $H>1/\alpha$, independence of the increments when $H=1/\alpha$, antipersistence of the increments when $H<1/\alpha$ and $\alpha$ not too small, and a last and newly observed regime when $H<1/\alpha$ and $\alpha$ small. In this last regime, there is a destruction of the memory like in the antipersistent case, but some past large increments are in some way unforgettable events, so that we observe a local persistence of the increments after a large increment. Finally, an application to real financial data underlines the relevance of the method.

A useful extension of our work would be to use the decomposition to simulate at discrete times an $\alpha$-stable process having the same codifference as an LFSM. It would thus only be an approximation of an LFSM but it would be interesting to compare it to other simulation methods, since no exact simulation method exists for the LFSM.

\bibliographystyle{plain}
\bibliography{cas-refs}

\appendix

\section{Proof of Proposition~\ref{pro:sumSaS}}\label{sec:proof_sumSaS}

\begin{proof}
Using the independence of the $X_i$ along with equation~\ref{eq:SaS}, we have, for any $\theta\in\mathbb R$,
$$\Phi_{\sum_{i=1}^d a_i X_i}(\theta) = \prod_{i=1}^d \Phi_{X_i}(a_i \theta)=\exp\left(-\sum_{i=1} ^d \Vert X_i\Vert _{\alpha} ^{\alpha}|a_i |^{\alpha}|\theta|^{\alpha}\right),$$
meaning that $\sum_{i=1}^{d}a_iX_i$ is $S\alpha S$ with scale parameter $\left(\sum_{i=1} ^d \Vert X_i\Vert _{\alpha} ^{\alpha}|a_i |^{\alpha}\right)^{1/\alpha}$. This leads to equation~\eqref{eq:sumSaS}.
\end{proof}

\section{Proof of Theorem~\ref{thm:systDecompo}}\label{sec:proof_systDecompo}

\begin{proof}
The autocodifference of the LFSM is provided in equation~\eqref{eq:codiffLFSM}. But the knowledge of the symmetric codifference matrix of $(X_t,...,X_{t+d-1})'$ is equivalent to the knowledge of $\Vert X_{t+i}\Vert_{\alpha}$, for $0\leq i\leq d-1$, and $\Vert X_{t+\ell}-X_{t+i}\Vert_{\alpha}$, for $0\leq i<\ell\leq d-1$. So we just look for coefficients $a_{t,i,j}$ such that $\Vert \mathcal T X_{t+i}\Vert_{\alpha}=\Vert X_{t+i}\Vert_{\alpha}$ and $\Vert \mathcal T X_{t+\ell}-\mathcal T X_{t+i}\Vert_{\alpha}=\Vert X_{t+\ell}-X_{t+i}\Vert_{\alpha}$. Using the independence and unitary scale of the $Z_j$ and Proposition~\ref{pro:sumSaS}, we easily get the expressions $\Vert \mathcal T X_{t+i}\Vert_{\alpha}^{\alpha}=\sum_{j=0}^{i} |a_{t,i,j}|^{\alpha}$ and $\Vert \mathcal T X_{t+\ell}-\mathcal T X_{t+i}\Vert_{\alpha}^{\alpha}=\sum_{j=0}^{\ell} |a_{t,\ell,j}-a_{t,i,j}|^{\alpha}$. Using the stationarity of the increments along with equation~\eqref{eq:scalParamLFSM}, we also get $\Vert X_{t+\ell}-X_{t+i}\Vert_{\alpha}^{\alpha}=K_{\alpha,H}^{\alpha}|\ell-i|^{\alpha H}$ and $\Vert X_{t+i}\Vert_{\alpha}^{\alpha}=K_{\alpha,H}^{\alpha}|t+i|^{\alpha H}$, so that we obtain the following system of $d(d+1)/2$ nonlinear equations:
$$\left\{\begin{array}{ll}
(e_i) & K_{\alpha,H}^{\alpha}|t+i|^{\alpha H} = \sum_{j=0}^{i} |a_{t,i,j}|^{\alpha} \\
(e_{\ell,i}) & K_{\alpha,H}^{\alpha}|\ell-i|^{\alpha H} = \sum_{j=0}^{\ell} |a_{t,\ell,j}-a_{t,i,j}|^{\alpha},
\end{array}\right.$$
with equation $(e_i)$ defined for $i\in\llbracket 0,d-1\rrbracket$ and equation $(e_{\ell,i})$ for $(i,\ell)\in\llbracket 0,d-2\rrbracket\times\llbracket i+1,d-1\rrbracket$. Thanks to an invertible transform of this system, we get the equivalent system of equations displayed in Theorem~\ref{thm:systDecompo}, with $(\mathcal E_{i,i})=(e_i)$ and $(\mathcal E_{\ell,i})=(e_{\ell})-(e_{\ell,i})$.

We now prove the uniqueness of the solution, assuming its existence. First, let's consider that $H=1/\alpha$. The condition $a_{t,\ell,j}=a_{t,i,j}$ when $j\leq i\leq \ell$ reduces the problem to the search of $d$ coefficients $a_{t,i,i}$, for $i\in\llbracket 0,d-1\rrbracket$, obtained by the $d$ equations $(\mathcal E_{i,i})$, which now write $|a_{t,i,i}|^{\alpha} = K_{\alpha,H}^{\alpha}(t+i) - \sum_{j=0}^{i-1} |a_{t,j,j}|^{\alpha}$. This linear problem can be written with a triangular matrix with nonzero diagonal coefficients, so the solution exists and is unique. Finally, noting that $K_{\alpha,H}=1$ when $H=1/\alpha$, we get the following expression for the coefficients: $a_{t,i,j}=t^{\frac{1}{\alpha}\indic_{j=0}}\indic_{i\geq j}$.

We now assume that $H\neq 1/\alpha$. We solve the system iteratively in the lexicographical order of the indices in $(\mathcal E_{\ell,i})$. There is no difficulty when $\ell=i$. When $\ell\neq i$, we have to prove that the function $f_{t,i}$ is injective. The function $f_{t,i}$ is differentiable in $(a_{t,i,i},+\infty)$, which is its domain of definition when $H>1/\alpha$, according to the condition $a_{t,\ell,i}>a_{t,i,i}$ given in the theorem. Its derivative is $f'_{t,i}(z)=\alpha(|z|^{\alpha-1}-|z-a_{t,i,i}|^{\alpha-1})$ in this interval. We note that when $H>1/\alpha$, then $\alpha>1$ since $H<1$. We also note that $|z-a_{t,i,i}|=z-a_{t,i,i}< z$. So $f'_{t,i}$ is strictly positive in $(a_{t,i,i},+\infty)$ and $f_{t,i}$ is injective.

The function $f_{t,i}$ is differentiable as well in $(0,a_{t,i,i})$, which is the relevant interval when $H<1/\alpha$, after the condition $a_{t,\ell,i}<a_{t,i,i}$. In this case, $f'_{t,i}(z)=\alpha(|z|^{\alpha-1}+|z-a_{t,i,i}|^{\alpha-1})$, which is again strictly positive in $(0,a_{t,i,i})$, so $f_{t,i}$ is injective.
\end{proof}

\section{Proof of Theorem~\ref{thm:projMetric}}\label{sec:proof_projMetric}

\begin{proof}
Since $\widehat{\mathcal T X}^{D,V_{0,d-2}}_{d}\in V_{0,d-2}$, we can write 
$$\widehat{\mathcal T X}^{D,V_{0,d-2}}_{d}= \sum_{j=0}^{d-2} b_{j}Z_j,$$
where $b_0$, ..., $b_{d-2}$ are to be determined. Replacing $\mathcal T X_d$ by its expression in equation~\eqref{eq:transfodiscr}, we are looking for the parameters $b_0$, ..., $b_{d-2}$ minimizing the (semi-)metric $(|a_{1,d-1,d-1}|^{\alpha}+\sum_{j=0}^{d-2}|b_j-a_{1,d-1,j}|^{\alpha})^{1/\alpha}$. Since it is a sum of positive terms, it is minimized when they are all (except $|a_{1,d-1,d-1}|^{\alpha}$) equal to zero, what happens iff $b_j=a_{1,d-1,j}$ for all $j\in\llbracket 0,d-2\rrbracket$.
\end{proof}

\end{document}